\documentclass[aps,prb,twocolumn,superscriptaddress,floatfix]{revtex4-1}
\usepackage{bbold}
\usepackage{mathrsfs}
\usepackage{graphicx,graphics}
\usepackage{dcolumn}
\usepackage{amsmath,amssymb,amsfonts}
\usepackage{latexsym,verbatim}
\usepackage{bm}

\usepackage{color}
\usepackage{ulem}
\usepackage{braket}
\usepackage[percent]{overpic}
\usepackage[breaklinks=true,colorlinks,citecolor=blue,linkcolor=blue,urlcolor=blue]{hyperref}

\begin{document}
\title{Theory of Photon Condensation in a Spatially-Varying Electromagnetic Field}
\author{G.M. Andolina}
\thanks{These two authors contributed equally.}
\affiliation{NEST, Scuola Normale Superiore, I-56126 Pisa,~Italy}
\affiliation{Istituto Italiano di Tecnologia, Graphene Labs, Via Morego 30, I-16163 Genova,~Italy}
\author{F.M.D. Pellegrino}
\thanks{These two authors contributed equally.}
\affiliation{Dipartimento di Fisica e Astronomia ``Ettore Majorana'', Universit\`a di Catania, Via S. Sofia 64, I-95123 Catania,~Italy}
\affiliation{INFN, Sez.~Catania, I-95123 Catania,~Italy}
\author{V. Giovannetti}
\affiliation{NEST, Scuola Normale Superiore and Istituto Nanoscienze-CNR, I-56126 Pisa,~Italy}
\author{A.H. MacDonald}
\affiliation{Department of Physics, The University of Texas at Austin, Austin, TX 78712, USA}
\author{M. Polini}
\affiliation{Dipartimento di Fisica dell'Universit\`a di Pisa, Largo Bruno Pontecorvo 3, I-56127 Pisa, Italy}
\affiliation{School of Physics \& Astronomy, University of Manchester, Oxford Road, Manchester M13 9PL, United Kingdom}
\affiliation{Istituto Italiano di Tecnologia, Graphene Labs, Via Morego 30, I-16163 Genova,~Italy}
\begin{abstract}
The realization of equilibrium superradiant quantum phases (photon condensates) in a spatially- uniform quantum cavity field is forbidden by a ``no-go" theorem stemming from gauge invariance. We here show that the no-go theorem does not apply to spatially-varying quantum cavity fields. We find a criterion for its occurrence that depends solely on the static, non-local orbital magnetic susceptibility $\chi_{\rm orb}(q)$, of the electronic system (ES) evaluated at a cavity photon momentum $\hbar q$. Only 3DESs satisfying the Condon inequality $\chi_{\rm orb}(q)>1/(4\pi)$ can harbor photon condensation. For the experimentally relevant case of two-dimensional (2D) ESs embedded in quasi-2D cavities the criterion again involves $\chi_{\rm orb}(q)$ but also the vertical size of the cavity. We use these considerations to identify electronic properties that are ideal for photon condensation. Our theory is non-perturbative in the strength of electron-electron interaction and therefore applicable to strongly correlated ESs.
\end{abstract}
\maketitle

\section{Introduction}
\label{sect:intro}

The Dicke model~\cite{dicke_pr_1954}, which describes a system of $N$ qubits coupled to a single-mode spatially-uniform field confined in a cavity of volume $V$, plays a central role in quantum optics and cavity quantum electrodynamics (QED)~\cite{gross_pr_1982,cong_josaB_2016,kockum_naturereviewsphysics_2019,kirton19}. In 1973 Hepp and Lieb~\cite{hepp_lieb} and subsequently Wang and Hioe~\cite{wang_pra_1973} pointed out
that for sufficiently strong light-matter coupling  the Dicke model in the thermodynamic limit ($N\to \infty$, $V\to \infty$, with $N/V={\rm const}$) has a finite temperature second-order equilibrium phase transition between a normal and ``superradiant" state. In the latter, the ground state contains a macroscopically large number of coherent 
photons, {\it i.e.}~$\braket{\hat{a}}\propto \sqrt{N}$, where $\hat{a}$ ($\hat{a}^\dagger$) destroys (creates) a cavity photon. To avoid confusion with the superradiant emission discussed in the original work by Dicke
we refer to the equilibrium superradiant phase as a photon condensate.
 Equilibrium superradiance was shown to be robust against the addition of counter-rotating 
terms~\cite{hepp_lieb_2,carmichael_physlett_1973} neglected in Refs.~\onlinecite{hepp_lieb,wang_pra_1973}, but not against restoration of an additional neglected term proportional to $(\hat{a}+\hat{a}^\dagger)^2$ (Ref.~\onlinecite{rzazewski_prl_1975}).  This quadratic term is naturally generated by applying minimal coupling 
$\hat{\bm p}\to \hat{\bm p}+e {\bm A}/c$ to the electron kinetic energy $\hat{\bm p}^2/(2m)$. Rza\.{z}ewski~et al.~\cite{rzazewski_prl_1975} were the first to show that the Thomas-Reiche-Kuhn (TRK) sum rule~\cite{Sakurai,Tufarelli15} poses an insurmountable obstacle against equilibrium superradiance in a spatially-uniform quantum cavity field. Physically, this sum rule originates from gauge invariance~\cite{Pines_and_Nozieres,Giuliani_and_Vignale}, and in particular from the property that a system cannot respond to a spatially-uniform and time-independent vector potential. The link between gauge invariance and quadratic terms emerges as following. The quadratic term is responsible for the appearance of a {\it diamagnetic} contribution to the current operator~\cite{Pines_and_Nozieres,Giuliani_and_Vignale}. Only when paramagnetic and diamagnetic contributions are considered on equal footing, does one have a precisely gauge-invariant Hamiltonian satisfying the TRK sum rule. Recent advances in technology have reinvigorated interest in equilibrium superradiance~\cite{emary_brandes,buzek_prl_2005}, inspiring a literature thread in which the obstacle presented by quadratic terms was periodically resurrected~\cite{rzazewski_prl_2006,nataf_naturecommun_2010}. Complications due to the presence of a superconducting condensate in circuit QED setups were also discussed~\cite{nataf_naturecommun_2010,viehmann_prl_2011,ciuti_prl_2012,Jaako16,Bamba16}. 

In the Dicke model direct interactions between two-level systems are neglected. Effective long-range interactions between qubits are solely mediated by the common cavity field. Recent experimental progress has created opportunities to study light-matter interactions in an entirely new
regime. For example, two-dimensional (2D) electron systems (ESs) can be embedded in cavities or exposed to the radiation field of 
metamaterials, making it possible to study strong light-matter interactions in the regime where direct electron-electron interactions may play a pivotal role, as in the quantum Hall regime~\cite{smolka_science_2014,pellegrino_natcom_2016,ravets_prl_2018,knuppel_nature_2019,scalari_science_2012,muravev_prb_2013,Paravicini-Bagliani_natphys_2019}. 

Similarly, one can imagine cavity QED in which matter exhibits strongly correlated phenomena~\cite{Schlawin_prl_2019,curtis_prl_2019,allocca_prb_2019,kiffner_prb_2019,raines_physrevreser_2020,Li20,Li20b,Ashida20,Rubio2018,Rubio2018a,Rubio2019,Rubio2020}
such as exciton condensation, superconductivity, magnetism, or Mott insulating states. For all these exciting new possibilities, the paradigmatic Dicke model needs of course to be transcended. The degrees of freedom of microscopic many-body Hamiltonians---such as the one of the jellium model~\cite{Giuliani_and_Vignale} or the Hubbard model~\cite{Hubbard} to name two---need to be coupled to the cavity modes. As the Dicke model story has instructed us, theories of the equilibrium properties of these intriguing new systems must be fully gauge invariant. This has not always been the case in the literature. For example, the case of materials with a low-energy linear energy-momentum dispersion relation, such as graphene and Weyl semimetals, is particularly tricky. In this case, the low-energy continuum model Hamiltonian needs to be accompanied by an ultraviolet cut-off, which breaks gauge invariance~\cite{abedinpour_prb_2011}. Using this model to study superradiant quantum phase transitions, e.g.~in graphene~\cite{hagenmuller_prl_2012},  incorrectly implies photon condensation because a dynamically generated quadratic term is missed~\cite{chirolli_prl_2012,pellegrino_prb_2014}. We therefore conclude that low-energy truncations of the Hilbert space must be carried out carefully in order to preserve gauge invariance~\cite{abedinpour_prb_2011,DiStefano19,Stokes20}. Another example is that of Ref.~\onlinecite{mazza_prl_2019}, where the coupling of the matter degrees of freedom of a two-band Hubbard model to the spatially-uniform vector potential of the cavity was carried out via a paramagnetic current operator not satisfying the continuity equation (see Ref.~\onlinecite{andolina_prb_2019} for further details). A no-go theorem for superradiant quantum phase transitions which is applicable to generic interacting many-body systems in a cavity has been recently demonstrated in Ref.~\onlinecite{andolina_prb_2019}, under the strong but almost universally made assumption of a spatially-uniform cavity field.

The term ``superradiance" is used to describe a plethora of different collective phenomena, ranging from the amplification of radiation due to coherence in the emitting medium~\cite{dicke_pr_1954} to the Zel'dovich-Misner-Unruh~\cite{Zel'dovich} amplification of radiation by rotating black holes. To avoid confusion, we will therefore refer to the equilibrium superradiant phase as a photon condensate. Given the impossibility of achieving photon condensation in a spatially-uniform quantum cavity field, in this Article we relax this strong assumption. We lay down a theory of photon condensation in a {\it spatially-varying} quantum cavity field that does not rely on the smallness of the electron-electron-interaction coupling constant. As such, our theory is applicable to strongly correlated ESs. For pioneering theoretical works on the case of spatially-varying quantum cavity fields see Refs.~\onlinecite{Gawedzki_pra_1981,Bamba_pra_2014,Basko19}.

We separately study three cases: 

i) We first consider a three-dimensional (3D) ES embedded in a 3D cavity field. In this case, we reach a condition for the occurrence of photon condensation which is universal, in that it does not depend on the cavity material parameters. Indeed, our criterion depends only on a non-local linear response function of the 3DES, namely the static non-local orbital magnetic susceptibility $\chi_{\rm orb}(q)$. This quantity describes the response of the electron system to a static but spatially-oscillating magnetic field:
\begin{equation}\label{eq:OMS}
\chi_{\rm orb}(q)\equiv -\frac{e^2}{c^2}\frac{\chi_{\rm T}(q,0)}{q^2}~.
\end{equation}
Here, $-e$ is the electron charge, $c$ is the speed of light in vacuum, and $\chi_{\rm T}(q,0)$ is the transverse current response function of the interacting ES~\cite{Pines_and_Nozieres,Giuliani_and_Vignale}. We find that photon condensation occurs if and only if $\chi_{\rm orb}(q)> 1/(4\pi)$. 

ii) We then study the role of spin degrees of freedom, by including in the treatment the Zeeman coupling between the electron spin and the spatially-varying cavity field. We also discuss the combined effects of orbital and spin couplings.

iii) Finally, we consider the case of a 2DES embedded in a quasi-2D cavity of extension $L_z$ in the direction perpendicular to the plane hosting the 2DES, i.e.~the $\hat{\bm x}$-$\hat{\bm y}$ plane~\cite{Basko19}. 
In this case, the criterion for photon condensation depends on $L_z$, and not only on the intrinsic orbital magnetic properties of the 2DES.

Our Article is organized as following. Photon condensation in 3D in the presence of purely orbital coupling between the cavity electromagnetic field and matter degrees of freedom is discussed in Sect.~\ref{sect:3D_stiffness_theorem}. The role of spin and combined orbital-spin effects (always in 3D) is reported in Sect.~\ref{sect:spin_and_orbital_3D_effects}. Finally, the case of 2DESs embedded in quasi-2D cavities is discussed in Sect.~\ref{sect:2D_stiffness_theorem}. A brief summary and our main conclusions are finally presented in Sect.~\ref{sect:summary_and_conclusions}. 
A number of cumbersome mathematical proofs and useful technical details are reported in Appendices~\ref{appendix:disentagling}-\ref{app:calculation_determinant}.

\section{3D Photon Condensation}
\label{sect:3D_stiffness_theorem}

We consider a 3DES interacting with a spatially-varying quantized electromagnetic field. For the sake of concreteness, we assume that the 3DES is described by the jellium model Hamiltonian~\cite{Pines_and_Nozieres,Giuliani_and_Vignale}
\begin{equation}\label{eq:jellium}
\hat{\cal H}=\sum_{i=1}^N  \frac{\hat{ {\bm p}}_i^2}{2m}+ \frac{1}{2}\sum_{i\neq j} v(|\hat{\bm r}_{i} - \hat{\bm r}_{j}|)~.
\end{equation}
This model describes $N$ electrons of mass $m$ interacting via an arbitrary~\cite{jellium} central potential $v(r)$. Charge neutrality (and therefore stability) of the system is guaranteed by a positive background of uniform charge. Electron-background and background-background interactions have  not been explicitly written in $\hat{\cal H}$. For future reference, we denote by $\ket{\psi_m}$ and $E_{m}$ the {\it exact} eigenstates and eigenvalues~\cite{Giuliani_and_Vignale,Pines_and_Nozieres} of 
$\hat{\cal H}$, with $\ket{\psi_{0}}$ and $E_{0}$ denoting the 
ground state and ground-state energy, respectively. We also introduce the 3D Fourier transforms of the density and paramagnetic (number) current operators~\cite{Giuliani_and_Vignale,Pines_and_Nozieres}:
\begin{eqnarray}
\label{eq:densityq}
\hat{n}(\bm{q})&=&{\sum_{i=1}^N e^{-i\bm{q}\cdot \hat{\bm{r}}_i}}~,\\
\hat{\bm{j}}_{\rm p}(\bm{q})&=&\frac{1}{2m} \sum_{i=1}^N \left(\hat{ {\bm p}}_i e^{-i\bm{q}\cdot \hat{\bm{r}}_i} + e^{-i\bm{q}\cdot \hat{\bm{r}}_i}\hat{ {\bm p}}_i\right)~,
\end{eqnarray}
with $\hat{n}(- \bm{q})= \hat{n}^\dagger(\bm{q})$ and $\hat{\bm{j}}_{\rm p}(- \bm{q})= \hat{\bm{j}}^\dagger_{\rm p}(\bm{q})$.

We treat the spatially-varying cavity electromagnetic field $\hat{{\bm A}}({\bm r})$ in a quantum fashion~\cite{grynberg,Walls_and_Milburn}. We consider a cavity of volume $V=L_{x}L_{y}L_{z}$, impose periodic boundary conditions on the cavity field, and represent it in terms of plane waves:
\begin{equation}\label{vectorpotential}
\hat{{\bm A}}(\bm{r})= \sum_{\bm{q},\sigma }  A_{\bm{q}} {\bm u}_{\bm{q},\sigma}(\hat{a}_{\bm{q},\sigma }e^{i\bm{q}\cdot \bm{r}} + \hat{a}_{\bm{q},\sigma}^\dagger  e^{-i\bm{q}\cdot \bm{r}})~.
\end{equation}
Here, ${\bm q}= (2\pi n_{x}/L_{x},2\pi n_{y}/L_{y}, 2\pi n_{z}/L_{z})$ with $(n_{x},n_{y},n_{z})$ relative integers, $\sigma=1,2$ is the polarization index, ${\bm u}_{\bm{q},\sigma}$ is the linear polarization vector,  $A_{\bm{q} }=\sqrt{2\pi \hbar c^2/(V \omega_{\bm{q}}\epsilon_{\rm r})}$, $\omega_{\bm q}=cq/\sqrt{\epsilon_{\rm r}}$, and $\epsilon_{\rm r}$ is a relative dielectric constant. The following properties hold~\cite{grynberg}: $\omega_{-{\bm q}}=\omega_{\bm q}$, ${\bm u}_{-{\bm q},\sigma}={\bm u}_{{\bm q},\sigma}$, $A_{-{\bm q}}=A_{\bm q}$, and ${\bm u}_{{\bm q},\sigma} \cdot 
{\bm u}_{{\bm q},\sigma^\prime}=\delta_{\sigma,\sigma^\prime}$. In the Coulomb gauge, we have the transversality condition 
\begin{equation}\label{eq:transv_condition}
{\bm u}_{{\bm q},\sigma}\cdot {\bm q} = 0~,
\end{equation}
for every ${\bm q}$ and $\sigma$.
The photonic annihilation and creation operators in Eq.~(\ref{vectorpotential}) satisfy bosonic commutation relations, $[\hat{a}_{{\bm q}, \sigma},\hat{a}^\dagger_{{\bm q}^\prime, \sigma^\prime}]=\delta_{{\bm q},{\bm q}^\prime}\delta_{\sigma,\sigma^\prime}$.

Being a quantum object, the field $\hat{{\bm A}}(\bm{r})$ has its own dynamics, which is determined by the photon Hamiltonian
\begin{equation}
\hat{\cal H}_{\rm ph}  = \sum_{{\bm q},\sigma}\hbar \omega_{\bm q} \left(\hat{a}^\dagger_{{\bm q}, \sigma}\hat{a}_{{\bm q}, \sigma} + \frac{1}{2}\right)~.
\end{equation}
The full Hamiltonian, including light-matter interactions, is therefore given by
\begin{equation}\label{eqHtot}
\hat{\cal H}_{{\bm A}}=\hat{\cal H}+\hat{\cal H}_{\rm ph} +\sum_{i=1}^{N} \frac{e}{ m c}   \hat{{\bm A}}(\bm{r}_i) \cdot \hat{ {\bm p}}_i 
+\sum_{i=1}^{N} \frac{e^2 }{2mc^2}\hat{{\bm A}}^2(\bm{r}_i)~.
\end{equation}
The third and fourth terms in Eq.~(\ref{eqHtot}) are often referred to respectively as the 
paramagnetic and diamagnetic contributions to the light-matter coupling Hamiltonian. 

With the aim of studying the potential existence of a quantum phase transition to a photon condensate and make therefore general statements about  the ground state $\ket{\Psi}$ of $\hat{\cal H}_{{\bm A}}$, the model (\ref{eqHtot}) must be extrapolated to the thermodynamic limit~\cite{hepp_lieb} $N\to\infty$, $V\to \infty$, with constant $N/V$. As shown in Appendix.~\ref{appendix:disentagling}, in this limit, $\ket{\Psi}$ does not contain light-matter entanglement, i.e.~we can take $\ket{\Psi}=\ket{\psi}\ket{\Phi}$, where $\ket{\psi}$ and $\ket{\Phi}$ are matter and light states.  We can therefore introduce the effective Hamiltonian for the photonic degrees of freedom, $\hat{\cal{H}}^{\rm eff}_{\rm ph}[{\psi}]\equiv   \braket{\psi| \hat{\cal H}_{{\bm A}} |\psi}$. Explicitly,

\begin{eqnarray}\label{eqHLAbb}
&\hat{\cal{H}}^{\rm eff}_{\rm ph}&[{\psi}]= \hat{\cal H}_{\rm ph} + \braket{ \psi|\hat{\cal H}|\psi} \nonumber \\
&+&
\sum_{\bm{q},\sigma }   \frac{e}{c} A_{\bm q}\left[ \hat{a}_{\bm{q},\sigma}  {\bm{j}}_{\rm p}(-\bm{q}   ) \cdot\bm{u}_{\bm{q},\sigma} +
 {\rm h.c.} \right] \nonumber\\ &+& \frac{e^2}{2mc^2}  \sum_{\bm{q},{\bm{q}^\prime},\sigma }  A_{\bm q}A_{\bm q^\prime}  \bm{u}_{\bm{q},\sigma}  \cdot\bm{u}_{\bm{q}^\prime,\sigma}   \times  \nonumber\\ &\times &  \left[ \hat{a}_{ \bm{q}^\prime,\sigma}^\dagger \hat{a}_{\bm{q},\sigma} n(\bm{q}^\prime-\bm{q} )+ 
  \hat{a}_{ \bm{q},\sigma} \hat{a}_{ \bm{q}^\prime,\sigma}^\dagger n(\bm{q} -\bm{q}^\prime)\right.+ \nonumber \\&+&  \left.\hat{a} _{ \bm{q},\sigma}\hat{a}_{ \bm{q}^\prime,\sigma}{ n(-\bm{q}
  -\bm{q}^\prime)} +\hat{a}_{\bm{q}^\prime,\sigma}^\dagger\hat{a}_{\bm{q},\sigma}^\dagger {n(\bm{q} +\bm{q}^\prime)} \right] ~.
\end{eqnarray}
where we have used the transversality condition in Eq.~(\ref{eq:transv_condition}), 
and introduced 
\begin{equation}\label{eq:nq}
 {n}(\bm{q})\equiv\braket{\psi|\hat{n}(\bm{q})|\psi}
\end{equation}
and 
\begin{equation}\label{eq:jq}
{\bm{j}}_{\rm p}(\bm{q})\equiv \braket{\psi|\hat{\bm{j}}_{\rm p}(\bm{q})|\psi}~.
\end{equation}

In the Coulomb gauge, 3D photon condensation is manifested by a non-zero value of the order parameter $\bar{\alpha}_{{\bm q},\sigma}\equiv\braket{\Phi|\hat{a}_{{\bm q},\sigma}|\Phi}$ emerging at a critical value of a suitable light-matter coupling constant~\cite{hepp_lieb,wang_pra_1973}.  At the quantum critical point (QCP), $\bar{\alpha}_{{\bm q},\sigma}$ is small. Note also that, near the QCP, the matter state can be written as $\ket{\bar{\psi}}=\ket{{\psi}_0}+\sum_{{\bm q},\sigma} \bar{\alpha}_{{\bm q},\sigma}\ket{{\delta\psi}_{{\bm q},\sigma}} +{\cal O}(\bar{\alpha}^2_{{\bm q},\sigma})$. Since the diamagnetic term in Eq.~(\ref{eqHLAbb}) is quadratic in $\bar{\alpha}_{{\bm q},\sigma}$, we can approximate the quantity $n(\bm{q})$ in the last two lines of this equation with its value in the absence of light-matter interactions, i.e.~we can safely take $n(\bm{q}) \simeq \braket{\psi_0|\hat{n}(\bm{q})|\psi_0}$. We now assume that the ground state $\ket{\Psi_{0}}$ of the 3DES in the absence of light-matter interactions is homogenous and isotropic, i.e.~$\braket{\psi_0|\hat{n}(\bm{q})|\psi_0}= N \delta_{{\bm q}, {\bm 0}}$. The reason why this assumption was made is obvious from the form of the diamagnetic term in Eq.~(\ref{eqHLAbb}): inhomogeneous ground states with $\braket{\psi_0|\hat{n}(\bm{q})|\psi_0}\neq N \delta_{{\bm q}, {\bm 0}}$ would couple modes with ${\bm q}\neq {\bm q}^\prime$, rapidly leading to a problem that is intractable with purely analytical methods. Under this assumption, the effective Hamiltonian reduces to:

\begin{eqnarray}
\label{eqHLAbb2}
&\hat{\cal{H}}^{\rm eff}_{\rm ph}&[{\psi}]=\braket{ \psi|\hat{\cal H}|\psi} + \nonumber \\ 
&+& \sum_{\bm{q},\sigma }   \frac{eA_{\bm q}}{c}  \left[ \hat{a}_{\bm{q},\sigma}  {\bm{j}}_{\rm p}(-\bm{q}   ) \cdot\bm{u}_{\bm{q},\sigma} +
 \hat{a}^\dagger_{\bm{q},\sigma}  {\bm{j}}_{\rm p}(\bm{q}   ) \cdot\bm{u}_{\bm{q},\sigma} \right]\nonumber \\  &+& \frac{1}{2}   \sum_{\bm{q},\sigma }
 \left[\hbar \widetilde{\omega}_{\bm q} + \hbar \widetilde{\omega}_{\bm q}\left( \hat{a}_{ \bm{q},\sigma} ^\dagger\hat{a}_{ \bm{q},\sigma}+\hat{a}_{ -\bm{q},\sigma} ^\dagger\hat{a}_{-\bm{q},\sigma} \right)\right.+\nonumber \\  &+& \left.2\Delta_{\bm q}  \left(\hat{a} _{ -\bm{q},\sigma}\hat{a}_{ \bm{q},\sigma}+\hat{a}_{\bm{q},\sigma}^\dagger\hat{a}_{-\bm{q},\sigma}^\dagger \right)\right] ~,
\end{eqnarray}
where $\Delta_{\bm{q}}\equiv Ne^2 A_{\bm q}^2/(2m c^2) $ with $\Delta_{\bm q}=\Delta_{-\bm q}$, and $\hbar \widetilde{\omega}_{\bm q}=\hbar {\omega_{\bm q}}+2\Delta_{\bm q}$. The term $\sum_{\bm{q},\sigma } \hbar \widetilde{\omega}_{\bm q}/2 $ is a vacuum contribution. Eq.~(\ref{eqHLAbb2}) is a quadratic function of the photonic operators and can be diagonalized via the following Bogoliubov transformation:
\begin{eqnarray}
\label{eq:modes1}
 \hat{a}_{\bm{q},\sigma}^\dagger&=&{\cosh(x_{\bm{q}})} \hat{b}^\dagger_{\bm{q},\sigma}-  {\sinh(x_{\bm{q}})}\hat{b}_{-\bm{q},\sigma}~,
\end{eqnarray}
where $\cosh(x_{\bm{q}})=(\lambda_{\bm{q}}+1)/(2\sqrt{\lambda_{\bm{q}}} )$, $\sinh(x_{\bm{q}})=(\lambda_{\bm{q}}-1)/(2\sqrt{\lambda_{\bm{q}}})$, and $\lambda_{\bm{q}}=\sqrt{1+ 4\Delta_{\bm{q}}/\hbar{\omega}_{\bm{q}}}$. In terms of the new bosonic operators $\hat{b}_{ \bm{q},\sigma} ^\dagger, \hat{b}_{ \bm{q},\sigma}$ the effective Hamiltonian reads as follows:
\begin{eqnarray}\label{eqHLAbb33}
&\hat{\cal{H}}^{\rm eff}_{\rm ph}&[{\psi}]=\braket{ \psi|\hat{\cal H}|\psi} +\sum_{\bm{q},\sigma }  
  \hbar  {\Omega}_{\bm q}   \left(\hat{b}_{ \bm{q},\sigma} ^\dagger\hat{b}_{ \bm{q},\sigma}  + \frac{1}{2} \right) + \nonumber \\  &+& 
\sum_{\bm{q},\sigma }  \frac{eA_{\bm q}}{ c\sqrt{\lambda_{\bm{q}}}}
\left[{\bm{j}}_{\rm p}(-\bm{q}   ) \cdot    \bm{u}_{\bm{q},\sigma}    \hat{b}_{\bm{q},\sigma}+{\rm H.c.}\right]~,
\end{eqnarray}
where $\hbar{\Omega}_{\bm q}=\hbar{\omega}_{\bm{q}}\lambda_{\bm q}$.

Being a sum of displaced harmonic oscillators, the ground state $\ket{\Phi}$ of $\hat{\cal{H}}^{\rm eff}_{\rm ph}[\psi]$, for every matter state $|\psi\rangle$, is a tensor product $\ket{\mathscr{B}} \equiv \otimes_{\bm{q},\sigma} \ket{\beta_{\bm{q},\sigma}}$ of coherent states of the $\hat{b}_{\bm{q},\sigma}$ operators\cite{Walls_and_Milburn,Serafini}, i.e.~$\hat{b}_{\bm{q}^\prime,\sigma^\prime} \ket{\mathscr{B}}=\beta_{\bm{q}^\prime,\sigma^\prime}   \ket{\mathscr{B}}$. Note that the order parameter ${\alpha}_{{\bm q},\sigma}$ introduced above is linearly-dependent on ${\beta}_{{\bm q},\sigma}$, i.e.~${\alpha}_{{\bm q},\sigma}={\cosh(x_{\bm{q}})} \beta^*_{\bm{q},\sigma}-  {\sinh(x_{\bm{q}})}\beta_{-\bm{q},\sigma}$. Hence, a non-zero ${\beta}_{{\bm q},\sigma}$ implies a non-zero ${\alpha}_{{\bm q},\sigma}$. From now on, we will therefore consider ${\beta}_{{\bm q},\sigma}$ as the order parameter, which can again be considered small at the QCP.

We now introduce the following energy functional, obtained by taking the expectation value of $\hat{\cal{H}}^{\rm eff}_{\rm ph}[\psi]$ over $ \ket{\mathscr{B}}$: $E[\{\beta_{\bm{q}, \sigma}\}, \psi]\equiv \braket{\Psi|\hat{\cal H}_{\bm A}|\Psi}=\braket{\mathscr{B}| \hat{\cal{H}}^{\rm eff}_{\rm ph}[\psi]|\mathscr{B}}$:

\begin{eqnarray}
\label{eqHLAd1}
E[\{\beta_{\bm{q}, \sigma}\}, \psi] &=&\braket{ \psi|\hat{\cal H}|\psi} + \sum_{\bm{q},\sigma } ~
  \hbar  {\Omega}_{\bm q} \left(|\beta_{ \bm{q},\sigma}|^2+\frac{1}{2}\right) \nonumber\\
  &+&\sum_{\bm{q},\sigma }   \frac{eA_{\bm q}}{ c\sqrt{\lambda_{\bm{q}}}} 
\left[   {\bm{j}}_{\rm p}(-\bm{q}   ) \cdot    \bm{u}_{\bm{q},\sigma}    \beta_{\bm{q},\sigma}+{\rm c.c.}\right]~.\nonumber\\
\end{eqnarray}
This needs to be minimized with respect to $\{\beta_{{\bm q},\sigma}\}$ and $|\psi\rangle$.  The minimization with respect to $\{\beta_{\bm{q},\sigma}\}$ can be done analytically by imposing the condition  $\partial_{\beta_{\bm{q},\sigma} ^*}E[\{\beta_{\bm{q},\sigma}\}, \psi]=0$. We find that the optimal value of  $\{\beta_{\bm{q},\sigma}\}$ is given by:
\begin{eqnarray}
\label{eq:min2}
\bar{\beta}_{\bm{q},\sigma}= -\frac{A_{\bm{q}}}{\hbar  \omega_{\bm{q}}\lambda^{3/2}_{\bm{q}}} \frac{e}{c}  {\bm{j}}_{\rm p}(\bm{q}) \cdot \bm{u}_{\bm{q},\sigma}~,
\end{eqnarray}
which depends on $\ket{\psi}$ through Eq.~(\ref{eq:jq}). Note that this equation can be written in terms of the operator
\begin{equation}\label{eq:operator B}
\hat{B}_{{\bm q}, \sigma} \equiv -\frac{A_{\bm{q}}}{\hbar  \omega_{\bm{q}} \lambda^{3/2}_{\bm{q}}} \frac{e}{c}  \hat{\bm{j}}_{\rm p}(\bm{q}) \cdot \bm{u}_{\bm{q},\sigma}~,
\end{equation}
i.e.~$\bar{\beta}_{\bm{q},\sigma}=\braket{\psi|\hat{B}_{{\bm q}, \sigma}|\psi}$.

Using Eq.~(\ref{eq:min2}) into Eq.~(\ref{eqHLAd1}), we finally find the energy functional that needs to be minimized with respect to $\ket{\psi}$:
\begin{equation}\label{eq:Emin}
E[\{\bar{\beta}_{\bm{q},\sigma}\}, \psi]=\braket{ \psi|\hat{\cal H}|\psi} - \sum_{\bm{q},\sigma }  \hbar  {\Omega}_{\bm q}\left( |\bar{\beta}_{ \bm{q},\sigma}|^2-\frac{1}{2}\right)~.
\end{equation}
As in the case of a spatially-uniform cavity field~\cite{andolina_prb_2019}, we are therefore left with a {\it constrained} minimum problem for the matter degrees of freedom: we need to seek the minimum of (\ref{eq:Emin}) among the normalized anti-symmetric states $\ket{\psi}$  which yield (\ref{eq:min2}). Such constrained minimum problems can be effectively handled with the stiffness theorem~\cite{Giuliani_and_Vignale}.

For photon condensation to occur we need the photon condensate phase to be energetically favored with respect to the normal phase, i.e.~we need $E[\{\bar{\beta}_{\bm{q},\sigma}\}, \psi]< E[0, \psi_0]$  or, equivalently,

\begin{eqnarray}
\label{eq:Emin11}
\braket{ \psi|\hat{\cal H}|\psi} -\braket{ \psi_0|\hat{\cal H}|\psi_0}< \sum_{\bm{q},\sigma }~
  \hbar  {\Omega}_{\bm q} |\bar{\beta}_{ \bm{q},\sigma}|^2~.
\end{eqnarray}
Note that the left-hand side of the previous inequality is the energy difference $E[\{\bar{\beta}_{\bm{q},\sigma}\}, \psi]-E[0, \psi_0]$,  so that the vacuum contribution $\sum_{\bm{q},\sigma } \hbar {\Omega}_{\bm q}  /2 $ drops out of the right-hand side.

The dependence of $\braket{ \psi|\hat{\cal H}|\psi}-\braket{ \psi_0|\hat{\cal H}|\psi_0}$ on $\bar{\beta}_{\bm{q},\sigma}$ can be calculated {\it exactly} up to order $\bar{\beta}_{ \bm{q},\sigma} ^2$ by using the stiffness theorem~\cite{Giuliani_and_Vignale}. 
The expansion of the left hand side of the inequality (\ref{eq:Emin11}) up to order $\bar{\beta}_{\bm{q},\sigma}^2$ is justified by the smallness of $\bar{\beta}_{\bm{q},\sigma}$ at the QCP. From now on, we exclude the trivial case $\braket{\psi_0|\hat{\bm{j}}_{\rm p}(\bm{q})|\psi_0}\neq 0$, requiring that $\braket{\psi_0|\hat{\bm{j}}_{\rm p}(\bm{q})|\psi_0}=0$ for all values of $\bm{q}$: for non-trivial photon condensate phases to occur, the ground state of the 3DES described by (\ref{eq:jellium}) is required to display no ground-state currents at all length scales.

Using the stiffness theorem~\cite{Giuliani_and_Vignale}, we find, up to second order in  $\bar{\beta}_{\bm{q},\sigma}$,
\begin{widetext}
\begin{equation}\label{eq:Stiffness}
\braket{ \psi|\hat{\cal H}|\psi}-\braket{ \psi_0|\hat{\cal H}|\psi_0} = -\frac{1}{2}  \sum_{\bm{q},\sigma }  \sum_{\bm{q}^\prime,\sigma^\prime }  \chi^{-1}_{\hat{B}_{\bm{q},\sigma},\hat{B}_{-\bm{q}^\prime,\sigma^\prime}}(0)\bar{ \beta}^*_{ \bm{q},\sigma} \bar{\beta}_{ \bm{q}^\prime,\sigma^\prime} ~,
\end{equation}
\end{widetext}
where  $\chi^{-1}_{\hat{B}_{\bm{q},\sigma},\hat{B}_{-\bm{q}^\prime,\sigma^\prime}}(0)$ is the inverse of the static response function $\chi_{\hat{B}_{\bm{q},\sigma},\hat{B}_{-\bm{q}^\prime,\sigma^\prime}}(0)$,  the operator $\hat{B}_{\bm{q},\sigma}$ has been introduced in Eq.~(\ref{eq:operator B}), and we have used the notation of Ref.~\onlinecite{Giuliani_and_Vignale}. 
Since the ground state of the 3DES has been taken to be homogenous and isotropic~\cite{Giuliani_and_Vignale},
\begin{equation}
\chi_{\hat{B}_{\bm{q},\sigma}, \hat{B}_{-\bm{q}^\prime,\sigma^\prime}}(0) =\chi_{\hat{B}_{\bm{q},\sigma},\hat{B}_{-\bm{q},\sigma}}(0)\delta_{\bm{q},\bm{q}^\prime} \delta_{\sigma,\sigma^\prime}~.
\end{equation}
As any other response function,  $\chi_{\hat{B}_{\bm{q},\sigma},\hat{B}_{-\bm{q},\sigma}}(0)$ has a Lehmann representation~\cite{Giuliani_and_Vignale,Pines_and_Nozieres} in terms of the exact eigenstates of the Hamiltonian (\ref{eq:jellium}),
\begin{widetext}
\begin{equation}\label{eq:exact_eigenstate_chiBB}
\chi_{\hat{B}_{\bm{q},\sigma},\hat{B}_{-\bm{q},\sigma}}(0) = -\frac{2 A^2_{\bm q}}{\hbar^2 \omega^2_{\bm q}\lambda^{3}_{\bm q} } \frac{e^2}{c^2} \sum_{n\neq 0}\frac{|\langle \psi_{n}|\hat{\bm j}_{\rm p}({\bm q})\cdot {\bm u}_{{\bm q}, \sigma}|\psi_{0}\rangle|^2}{E_{n}-E_{0}}< 0~.
\end{equation}
We readily recognize $\chi_{\hat{B}_{\bm{q},\sigma},\hat{B}_{-\bm{q},\sigma}}(0)$ to be intimately linked to the static, paramagnetic current-current response tensor~\cite{Giuliani_and_Vignale}
\begin{equation}
\chi_{\hat{j}_{{\bm p}, i}({\bm q}),\hat{j}_{{\bm p}, k}(-{\bm q})}(0) = 
- \frac{1}{V}\sum_{n\neq 0}\frac{\braket{\psi_{n}|\hat{j}_{{\rm p}, i}({\bm q})|\psi_{0}}\braket{\psi_{0}|\hat{j}_{{\rm p}, k}(-{\bm q})|\psi_{n}} }{E_{n}-E_{0}} - \frac{1}{V}\sum_{n\neq 0}\frac{\braket{\psi_{0}|\hat{j}_{{\rm p}, i}({\bm q})|\psi_{n}}\braket{\psi_{n}|\hat{j}_{{\rm p}, k}(-{\bm q})|\psi_{0}} }{E_{n}-E_{0}}~,
\end{equation}
where $\hat{j}_{{\rm p}, i}({\bm q})$, with $i=x,y,z$, denotes the $i$-th Cartesian component of $\hat{\bm j}_{\rm p}({\bm q})$. Indeed, it is easy to show that
\begin{equation}\label{eq:chi_BB_interms_of_paramagnetic_tensor}
\chi_{\hat{B}_{\bm{q},\sigma},\hat{B}_{-\bm{q},\sigma}}(0) = \frac{A^2_{\bm q} N}{\hbar^2 \omega^2_{\bm q}\lambda^{3}_{\bm q} n} \frac{e^2}{c^2}\sum_{i,k}u^{(i)}_{{\bm q},\sigma}u^{(k)}_{{\bm q},\sigma}\chi_{\hat{j}_{{\bm p}, i}({\bm q}),\hat{j}_{{\bm p}, k}(-{\bm q})}(0)~,
\end{equation}
\end{widetext}
where $u^{(i)}_{{\bm q},\sigma}$ denotes the $i$-th Cartesian component of the vector 
${\bm u}_{{\bm q}, \sigma}$ and we have introduced the electron density $n=N/V$. The previous result can be written in a more transparent manner by introducing the {\it  physical}~ current-current response tensor~\cite{Giuliani_and_Vignale}, which contains a diamagnetic as well as a paramagnetic contribution:
\begin{eqnarray}\label{eq:chi_phys}
\chi^{\rm J}_{i,k}(\bm{q},0)=\frac{n}{m}\delta_{i,k} +\chi_{\hat{j}_{{\rm p}, i}(\bm{q}),\hat{j}_{{\rm p}, k}(-\bm{q})}(0)~. \nonumber \\
\end{eqnarray}
In a homogeneous and isotropic system, the rank-$2$ tensor $\chi^{\rm J}_{i,k} (\bm{q},0)$ can be decomposed in terms of the longitudinal and transverse current-current response functions~\cite{Giuliani_and_Vignale}, $\chi^{\rm J}_{\rm L}(q,0)$  and $\chi^{\rm J}_{\rm T}(q,0)$, respectively:
\begin{equation}\label{eq:decomposition long and trans}
\chi^{\rm J}_{i,k}(\bm{q},0) =\chi^{\rm J}_{\rm L}(q,0)\frac{q_i q_k}{q^2}+\chi^{\rm J}_{\rm T}(q,0)\left(\delta_{i,k} -\frac{q_i q_k}{q^2}\right)~.
\end{equation}
Note that, as a consequence of gauge invariance, $\chi^{\rm J}_{\rm L}(q,0)=0$ for every $q$~\cite{Giuliani_and_Vignale}.
Using Eqs.~(\ref{eq:chi_phys})-(\ref{eq:decomposition long and trans}) in Eq.~(\ref{eq:chi_BB_interms_of_paramagnetic_tensor}), we finally find
\begin{equation}\label{eq:chi_BB_interms_of_transverse_current_response1}
\chi_{\hat{B}_{\bm{q},\sigma},\hat{B}_{-\bm{q},\sigma}}(0) = \frac{A^2_{\bm q} N}{\hbar^2 \omega^2_{\bm q}\lambda^{3}_{\bm q} n} \frac{e^2}{c^2}\left[\chi^{\rm J}_{\rm T}(q,0)-\frac{n}{m}\right]~.
\end{equation}
As a natural consequence of the transversality of the electromagnetic field, imposed by the Coulomb gauge, only the transverse current-current response function $\chi^{\rm J}_{\rm T}(q,0)$ enters Eq.~\eqref{eq:chi_BB_interms_of_transverse_current_response1}.

We now return to the result of the stiffness theorem. Inserting Eq.~(\ref{eq:Stiffness}) inside Eq.~\eqref{eq:Emin11}, we finally find the condition for photon condensation in a 3DES embedded in a spatially-varying electromagnetic field:
\begin{eqnarray}\label{eq:Stiffness1}
- \sum_{\bm{q},\sigma } \left[  \frac{1}{ 2\chi_{\hat{B}_{\bm{q},\sigma},\hat{B}_{-\bm{q} ,\sigma}}(0)}+  \hbar  {\Omega}_{\bm q}  \right] |\bar{\beta}_{ \bm{q},\sigma}|^2< 0~.
\end{eqnarray}

Since we want to minimize the energy difference $E[\{\bar{\beta}_{\bm{q},\sigma}\}, \psi]- E[0, \psi_0]$, the optimal choice of $\bar{\beta}_{ \bm{q}\sigma}$ is constructed as follows: i) modes with momentum $\bm{q}$ and polarization $\sigma$ such that Eq.~\eqref{eq:Stiffness1} is satisfied acquire a finite displacement $\bar{\beta}_{ \bm{q}\sigma}\neq 0$, since this choice lowers the energy difference; ii) on the other hand, modes for which Eq.~\eqref{eq:Stiffness1} is not satisfied, are forced to be unpopulated, i.e.~to have $\bar{\beta}_{ \bm{q},\sigma}=0$. A finite occupation of these modes would indeed increase the energy difference. Hence, we can analyze the inequality (\ref{eq:Stiffness1}) for a fixed $\bm{q}$:
\begin{eqnarray}
\label{eq:Stiffness2}
- \chi_{\hat{B}_{\bm{q},\sigma},\hat{B}_{-\bm{q},\sigma }}(0) >\frac{1}{2 \hbar  {\Omega}_{\bm q}}~.
\end{eqnarray} 
Using Eq.~\eqref{eq:chi_BB_interms_of_transverse_current_response1} and the microscopic expression of $A_{\bm q}$, we can rewrite Eq.~\eqref{eq:Stiffness2} as follows:
\begin{equation}\label{eq:Stiffness4}
-4\pi \frac{c^2}{\omega^2_{\bm q}\epsilon_{\rm r}}\frac{e^2}{c^2} \left[\chi^{\rm J}_{\rm T}(q,0)-\frac{n}{m}\right]>1+ 4 \frac{\Delta_{\bm q}}{\hbar \omega_{\bm q}}~. 
\end{equation}

Before further simplifying Eq.~(\ref{eq:Stiffness4}), we wish to make a few observations on the special case of a single-mode spatially-uniform field: 
\begin{itemize}
\item[i)] {\it No-go theorem in the presence of the diamagnetic term}. Let us consider the standard situation in the literature, in which matter degrees of freedom are minimally coupled to a quantum field, which is assumed to be single mode and spatially uniform, with angular frequency $\omega_{0}$ and amplitude $A_{\bm q}=A_{0} = \sqrt{2\pi \hbar c^2/(V \omega_{0} \epsilon_{\rm r})}$. Consistently, if the assumption of spatial uniformity is done from the very beginning, by setting ${\bm q}={\bm 0}$ in Eq.~(\ref{vectorpotential}), one has to replace $\chi^{\rm J}_{\rm T}(q,0)$ with $\lim_{q\to 0} \chi^{\rm J}_{\rm T}(q,0)$ inside the square bracket in Eq.~(\ref{eq:Stiffness4}). In systems with no long-range order (i.e.~in systems that do not become superconducting), it is well known~\cite{Giuliani_and_Vignale} that the ``diamagnetic sum rule" holds true: $\lim_{q\to 0} \chi^{\rm J}_{\rm T}(q,0)=0$. In this case, Eq.~(\ref{eq:Stiffness4}) reduces to:
\begin{equation}
\label{eq:Stiffness4_uniform}
4\pi \frac{c^2}{\omega^2_{0}\epsilon_{\rm r}}\frac{e^2}{c^2}\frac{n}{m}>1+ 4 \frac{\Delta_{0}}{\hbar \omega_{0}}~,
\end{equation}
with $\Delta_{0} = e^2 N A^2_{0}/(2 m c^2)$. The left-hand-side of Eq.~(\ref{eq:Stiffness4_uniform}) can be easily seen to be equal to $4 \Delta_{0}/(\hbar \omega_{0})$ and this inequality therefore reduces to $0>1$, which is clearly absurd. This is the no-go theorem~\cite{andolina_prb_2019} for photon condensation in a single-mode spatially-uniform quantum field.

\item[ii)] {\it Spurious ``go theorem" in the absence of the diamagnetic term.} Neglecting artificially the diamagnetic contribution to Eq.~(\ref{eqHtot}) is equivalent to setting $\Delta_{0}=0$ in the right-hand-side of Eq.~(\ref{eq:Stiffness4_uniform}). In this case a photon condensate occurs provided that the Drude weight ${\cal D}=\pi e^2n/m$ of the 3DES satisfies the following inequality:
\begin{equation}
\label{eq:gross_mistake}
{\cal D}>\frac{\omega^2_{0}\epsilon_{\rm r}}{4}~.
\end{equation}
\end{itemize}

Returning to Eq.~(\ref{eq:Stiffness4}) and using in it the microscopic expressions for $\omega_{\bm q}$ and $\Delta_{\bm q}$ given above (right after Eq.~(\ref{vectorpotential}) and Eq.~(\ref{eqHLAbb2}), respectively), we finally conclude that a photon condensate phase occurs if and only if the following inequality is satisfied:
\begin{eqnarray}
\label{eq:Stiffness6}
-\frac{e^2}{c^2}\frac{\chi^{\rm J}_{\rm T}(q,0)}{q^2}    >\frac{1}{4\pi}~.
\end{eqnarray} 
The left-hand-side of Eq.~\eqref{eq:Stiffness6} has a very clear physical interpretation. It is the non-local orbital magnetic susceptibility~\cite{Giuliani_and_Vignale}
\begin{eqnarray}\label{eq:Orbit}
\chi_{\rm orb}(q)\equiv-\frac{e^2}{c^2} \frac{\chi^{\rm J}_{\rm T}(q,0)}{q^2}~,
\end{eqnarray} 
which, in the long-wavelength $q\to 0$ limit, reduces to the thermodynamic (i.e.~macroscopic) orbital magnetic susceptibility (OMS)
\begin{equation}\label{eq:OMS_3D}
\chi_{\rm OMS}\equiv \lim_{q\to 0}\chi_{\rm orb}(q)=\left.\frac{\partial M_{\rm O}}{\partial B}\right|_{B=0}~.
\end{equation}
Here, $M_{\rm O}$ is the orbital contribution to the magnetization. This limit exists in systems with no long-range order: indeed, $\chi^{\rm J}_{\rm T}(q,0)$ vanishes like $q^2$ in the long-wavelength $q\to 0$ limit, in agreement with the diamagnetic sum rule~\cite{Giuliani_and_Vignale}.

In summary, introducing $\chi_{\rm orb}(q)$, we can write Eq.~(\ref{eq:Stiffness6}) as
\begin{eqnarray}\label{eq:Stiffness7}
\boxed{\chi_{\rm orb}(q) >\frac{1}{4\pi}}~.
\end{eqnarray} 
Eq.~(\ref{eq:Stiffness7}) is the most important result of this Section, representing a rigorous criterion for the occurrence of photon condensation in a 3DES. 

\subsection{Discussion}
\label{sect:3D_discussion}
A few comments are now in order.

i) In 3D, as clear from Eq.~(\ref{eq:Stiffness7}), $\chi_{\rm orb}(q)$ is dimensionless. It therefore naturally plays the role of a coupling constant determining the strength of light-matter interactions. Only when it exceeds the value $1/(4\pi) \sim 0.08$ can photon condensation take place. 

ii) The criterion (\ref{eq:Stiffness7}) does not depend {\it explicitly} on $\epsilon_{\rm r}$ but only implicitly, through the $\epsilon_{\rm r}$-dependence of the e-e interaction potential~\cite{jellium} $v(r)$. The latter, in turn, has an impact on $\chi_{\rm orb}(q)$. 

iii) Note that, while $\chi_{\hat{B}_{\bm{q},\sigma},\hat{B}_{-\bm{q},\sigma}}(0)$ in Eq.~(\ref{eq:exact_eigenstate_chiBB}) and~(\ref{eq:chi_BB_interms_of_transverse_current_response1}) is negative definite, the transverse contribution $\chi_{\rm T}(q,0)$ to the current-current response function satisfies the inequality $\chi_{\rm T}(q,0)<n/m$ and can therefore be both positive or negative. In turn, this implies that, for a given 3DES, $\chi_{\rm OMS}$ can be positive or negative (and perhaps change sign with microscopic parameters such as the electron density $n$). Broadly speaking, materials can be divided intro two groups, from the point of view of their orbital response: a) orbital {\it diamagnets}, those which have $\chi_{\rm OMS}<0$, are most common. They will not display photon condensation, according to our criterion (\ref{eq:Stiffness7}); b) orbital {\it paramagnets}, those for which $\chi_{\rm OMS}>0$, are much more rare in nature but, as discussed below, do exist. Only orbital paramagnets with $\chi_{\rm OMS}>1/(4\pi)$ can display photon condensation. 

Just as an example, we remind the reader that for free (i.e.~non-interacting) parabolic-band fermions in 3D~\cite{Giuliani_and_Vignale},
\begin{equation}\label{eq:Landau_diamagnetism}
\chi^{(0)}_{\rm OMS}= - \frac{\alpha^2}{r_{\rm s}}\left(\frac{1}{768 \pi^5}\right)^{1/3}<0~,
\end{equation}
where $r_{\rm s} = [3/(4\pi n a^3_{\rm B})]^{1/3}$ is the so-called Wigner-Seitz or gas parameter, $a_{\rm B}=\hbar^2/(m e^2)$ is the Bohr radius, and $\alpha=e^2/(\hbar c)$ is the fine structure constant.

iv) The result in Eq.~\eqref{eq:Stiffness7} can be understood as the condition for the occurrence of a static magnetic instability~\cite{Basko19}. 
Indeed, let us consider the energy functional of a material subject to a magnetic field $\bm{H}(\bm{r})$: 
\begin{equation}
E[{\bm B}(\bm{r})]=\frac{1}{2} \int d^3\bm{r}~\bm{H}(\bm{r})\cdot  \bm{B}(\bm{r})~,
\end{equation} 
where $\bm{B}(\bm{r})$ is the magnetic induction. The latter is related to the magnetic field via the orbital magnetization $\bm{M}(\bm{r})$, i.e.~$\bm{B}(\bm{r})=\bm{H}(\bm{r})+4\pi \bm{M}(\bm{r})$. The difference between $\bm{H}$ and $\bm{B}$ stems from the flow of charges in response to ${\bm H}$, which creates an orbital magnetization ${\bm M}$. In the realm of linear response theory, we can relate the orbital magnetization to the magnetic induction, $\bm{M}(\bm{r})=\int d^3 \bm{r}^\prime \chi_{\rm orb}(|\bm{r}-\bm{r}^\prime|)\bm{B}(\bm{r}^\prime)$. We can therefore write the energy as a quadratic function of  $\bm{B}(\bm{r})$:
\begin{eqnarray}
E[{\bm B}(\bm{r})]&=&\frac{1}{2} \int d^3\bm{r} \int d^3\bm{r}^\prime \Big[\delta({\bm r}-{\bm r}^\prime)  \nonumber\\
&-&4\pi \chi_{\rm orb}(|\bm{r}-\bm{r}^\prime|)\Big]  \bm{B}(\bm{r}^\prime)\cdot  \bm{B}(\bm{r})~.
\end{eqnarray}
An instability occurs if $E[{\bm B}(\bm{r})]<0$, i.e.~if and only if $\bm{B}(\bm{r}) < 4\pi \int  d \bm{r}^\prime      \chi_{\rm orb}(|\bm{r}-\bm{r}^\prime|)\bm{B}(\bm{r}^\prime)$. Fourier transforming with respect to ${\bm r}$ yields Eq.~\eqref{eq:Stiffness7}. 

Magnetostatic instabilities and the criterion \eqref{eq:Stiffness7} have been discussed long ago~\cite{condon_physrev_1966,ying_prb_1970,Shoenberg,gordon_prl_1998,solta_physicaB_2002,logoboy_physicaB_2008}. In a 3D metal, the de Haas-van Alphen effect (oscillations of the magnetization in response to an applied magnetic field) can lead to a thermodynamic instability of the electron gas. The magnetization is a function of the magnetic induction and when the orbital magnetic susceptibility $\chi_{\rm OMS}$ obeys the inequality \eqref{eq:Stiffness7}, the magnetic induction is a multi-valued function of the field. Condon first pointed out that Maxwell's construction yields phase coexistence and the formation of  (paramagnetic and diamagnetic) domains. These ``Condon domains", although first predicted for Be~\cite{condon_physrev_1966}, were first unambiguously observed in Ag~\cite{condon_prl_1968}. Since then, Condon domains have been observed also in Be~\cite{solt_prl_1996}, Sn~\cite{solt_prb_2000}, and also Al, Pb, and In (for a recent review see, for example, Ref.~\onlinecite{solta_physicaB_2002}). They have also been observed in ${\rm Br}_2$-intercalated graphite~\cite{markiewicz_prl_1985}, which is a layered compound with quasi-2D character.

The derivation in Sect.~\ref{sect:3D_stiffness_theorem} shows that 3D photon condensation and Condon domain formation are the {\it same} phenomenon~\cite{Basko19}. 
In essence, the proof reported in Sect.~\ref{sect:3D_stiffness_theorem} is a fully quantum mechanical derivation of the condition for the occurrence of Condon domains, which transcends the usual semiclassical approximations~\cite{Shoenberg} used to derive (\ref{eq:Stiffness7}).

v) For the remainder of this Article (particularly for Sect.~\ref{sect:2D_stiffness_theorem}), it is useful to derive Eq.~(\ref{eqHLAbb2}) in an alternative way. 

Instead of determining the exact photonic state, as we did above, we now follow a much more humble approach. We evaluate the expectation value of the Hamiltonian \eqref{eqHLAbb2} on a trial photonic wavefunction of the form $\ket{\mathscr{A}} \equiv \otimes_{\bm{q},\sigma} \ket{\alpha_{\bm{q},\sigma}}$, namely a tensor product of coherent states of the $\hat{a}_{\bm{q},\sigma}$ operators, i.e.~$\hat{a}_{\bm{q}^\prime,\sigma^\prime} \ket{\mathscr{A}}=\alpha_{\bm{q}^\prime,\sigma^\prime} \ket{\mathscr{A}}$. (We know that the exact eigenstate is not of this form, i.e.~it is a tensor product $\ket{\mathscr{B}} \equiv \otimes_{\bm{q},\sigma} \ket{\beta_{\bm{q},\sigma}}$ of coherent states of the $\hat{b}_{\bm{q},\sigma}$ operators. Momentarily, we will understand what error is made in using $\ket{\mathscr{A}}$ rather than  $\ket{\mathscr{B}}$.) Such expectation value is easily obtained by replacing the photonic operators  in Eq.~\eqref{eqHLAbb2} with $c$-numbers, i.e.~by replacing $ \hat{a}_{\bm{q},\sigma}\to  \alpha_{\bm{q},\sigma}$. Up to a constant factor, we find
\begin{eqnarray}
\label{eqHLAbbNumber}
&\tilde{E}&[\{\alpha_{\bm{q}, \sigma}\}, \psi] 
\equiv \braket{\mathscr{A}| \hat{\cal{H}}^{\rm eff}_{\rm ph}[\psi]|\mathscr{A}} = \braket{ \psi|\hat{\cal H}|\psi} + \nonumber \\ 
&+& \sum_{\bm{q},\sigma }   \frac{eA_{\bm q}}{c}  \left[ \alpha_{\bm{q},\sigma}  {\bm{j}}_{\rm p}(-\bm{q}   ) \cdot\bm{u}_{\bm{q},\sigma} +
\alpha^*_{\bm{q},\sigma}  {\bm{j}}_{\rm p}(\bm{q}   ) \cdot\bm{u}_{\bm{q},\sigma} \right]\nonumber \\  &+& \frac{1}{2}   \sum_{\bm{q},\sigma }
 \left[ \hbar \widetilde{\omega}_{\bm q}\left(\alpha_{ \bm{q},\sigma} ^* \alpha_{ \bm{q},\sigma}+\alpha_{ -\bm{q},\sigma} ^*\alpha_{-\bm{q},\sigma} + 1 \right)\right.+\nonumber \\  &+& \left.2\Delta_{\bm q}  \left(\alpha_{ -\bm{q},\sigma}\alpha_{ \bm{q},\sigma}+\alpha_{\bm{q},\sigma}^*\alpha_{-\bm{q},\sigma}^* \right)\right]~.
\end{eqnarray}
Performing in Eq.~\eqref{eqHLAbbNumber}  the linear transformation $\alpha_{\bm{q},\sigma}^*={\cosh(x_{\bm{q}})} \beta^*_{\bm{q},\sigma}- {\sinh(x_{\bm{q}})}\beta_{-\bm{q},\sigma}$, analogous to Eq.~\eqref{eq:modes1}, we get:
\begin{eqnarray}
\label{eqHLAd1bis}
\tilde{E}[\{\beta_{\bm{q}, \sigma}\}, \psi] &=&\braket{ \psi|\hat{\cal H}|\psi} + \sum_{\bm{q},\sigma } ~\left(\frac{  \hbar \widetilde{\omega}_{\bm q} }{2}+
  \hbar  {\Omega}_{\bm q} |\beta_{ \bm{q},\sigma}|^2\right)\nonumber\\
  &+&\sum_{\bm{q},\sigma }   \frac{eA_{\bm q}}{ c\sqrt{\lambda_{\bm{q}}}} 
\left[   {\bm{j}}_{\rm p}(-\bm{q}   ) \cdot    \bm{u}_{\bm{q},\sigma}    \beta_{\bm{q},\sigma}+{\rm c.c.}\right]~.\nonumber\\
\end{eqnarray}
The quantity $\tilde{E}[\{\beta_{\bm{q}, \sigma}\}, \psi]$ differs from the exact result in Eq.~\eqref{eqHLAd1} only for the vacuum contribution, which is $\sum_{\bm{q},\sigma }  \hbar \widetilde{\omega}_{\bm q} /{2}$ instead of the correct one $ \sum_{\bm{q},\sigma }  \hbar  {\Omega}_{\bm q}/{2}$. However, since we are interested only in energy differences, the vacuum contribution drops out of the problem and the two procedures yield the same energy difference: $E[\{\bar{\beta}_{\bm{q},\sigma}\}, \psi]-E[0, \psi_0] =  \tilde{E}[\{\bar{\beta}_{\bm{q},\sigma}\}, \psi]-E[0, \psi_0]$. 

In conclusion, if one is solely interested in energy differences, it is not necessary to determine the eigenstates exactly but it is sufficient to assume the photonic wave-function to be a tensor product of coherent states of the $\hat{a}_{\bm{q},\sigma}$ operators.

\section{The role of Zeeman coupling and combined orbital-spin effects}
\label{sect:spin_and_orbital_3D_effects}

In this Section we investigate the role of the Zeeman coupling. To begin with, we consider (Sect.~\ref{sect:Zeeman_alone}) the case in which the 3DES couples to the radiation field only via the Zeeman term. In the second part of this Section (Sect.~\ref{sect:combined_orbital_Zeeman}), we consider the combined role of orbital and Zeeman couplings. The derivation of the corresponding criteria for photon condensation closely follows the case of pure orbital coupling discussed in Sect.~\ref{sect:3D_stiffness_theorem}.

\subsection{Light-matter interactions via the Zeeman term}
\label{sect:Zeeman_alone}

If the 3DES couples to the spatially-varying cavity electromagnetic field only via the Zeeman term, the full Hamiltonian is:
\begin{equation}\label{eqHtot9}
\hat{\cal H}_{{\bm B}}=\hat{\cal H} + \hat{\cal H}_{\rm ph} + \frac{g\mu_{\rm B}}{2}\sum_{i=1}^N \hat{\bm{\sigma}}_i\cdot \hat{\bm{B}}(\bm{r}_i)~,
\end{equation}
where $g$ is the Land\'e $g$-factor, $\mu_{\rm B}$ is the Bohr magneton, $\hat{\bm{\sigma}}_i$ is the spin operator of the $i$-th electron, and $\hat{\bm{B}}(\bm{r})=\bm{\nabla} \times \hat{\bm{A}}(\bm{r})$ is the magnetic component of the cavity electromagnetic field, $\hat{\bm{A}}(\bm{r})$ being given in Eq.~\eqref{vectorpotential}. Explicitly, the magnetic field reads as follows:
\begin{equation}
\hat{\bm{B}}(\bm{r})=\sum_{\bm{q},\sigma }  i q A_{\bm{q}}\bm{u}_{{\rm T},{\bm{q},\sigma}} \big( \hat{a}_{\bm{q} }e^{i\bm{q}\cdot \bm{r}} -     \hat{a}^\dagger_{\bm{q} }e^{-i\bm{q}\cdot \bm{r}}\big)~,
\end{equation}
where $\bm{u}_{{\rm T},{\bm{q},\sigma}}\equiv (\bm{q}/q)\times \bm{u}_{\bm{q},\sigma}$. (Note that $\{\bm{q}, \bm{u}_{\bm{q},\sigma}, \bm{u}_{{\rm T},{\bm{q},\sigma}}\}$ is a set of orthogonal vectors.)

As shown in Appendix~\ref{appendix:disentagling2}, the ground state $\ket{\Psi}$ of $\hat{\cal H}_{{\bm B}}$ does not contain light-matter
entanglement in the thermodynamic limit, i.e.~we can take $\ket{\Psi}=\ket{\psi}\ket{\Phi}$, where $\ket{\psi}$
and  $\ket{\Phi}$ are matter and light states. As in Sect.~\ref{sect:3D_stiffness_theorem}, we are therefore led to introduce an effective Hamiltonian for the photonic degrees of freedom, $\hat{\cal{H}}^{\rm eff}_{\rm ph}[{\psi}]\equiv \braket{\psi| \hat{\cal H}_{{\bm B}} |\psi}$:

\begin{eqnarray}\label{eqHeffSpin}
\hat{\cal{H}}^{\rm eff}_{\rm ph}[{\psi}]&=&\braket{\psi|\hat{\cal H}|\psi}+\hat{\cal H}_{\rm ph} +\nonumber \\&+& \sum_{\bm{q},\sigma } \frac{g\mu_{\rm B}  A_{\bm{q}} }{2} \big[{\bm{S}}  (-\bm{q})   \hat{a}_{\bm{q},\sigma}  -   {\bm{S}}  (\bm{q})  \hat{a}^\dagger_{\bm{q},\sigma}\big]\cdot     iq\bm{u}_{{\rm T},{\bm{q},\sigma}} ~,\nonumber \\
\end{eqnarray}
where~\cite{Giuliani_and_Vignale}
\begin{equation}
\hat{ \bm{S}}(\bm{q})=\sum_{i=1}^Ne^{-i\bm{q}\cdot \bm{r}_i}\hat{\bm{\sigma}}_i
\end{equation}
is the 3D Fourier transform of the spin density $\hat{\bm{S}}(\bm{q}) = \sum_{i=1}^N \hat{\bm{\sigma}}_i \delta({\bm r}-{\bm r}_i)$ and $\bm{S}  (\bm{q})=\braket{\psi|\hat{ \bm{S}}  (\bm{q})|\psi}$.

Since Eq.~\eqref{eqHeffSpin} is a sum of displaced harmonic oscillators, we can assume without loss of generality that the ground state $\ket{\Phi}$ of $\hat{\cal{H}}^{\rm eff}_{\rm ph}[\psi]$ is a tensor product $\ket{\mathscr{A}} \equiv \otimes_{\bm{q},\sigma} \ket{\alpha_{\bm{q},\sigma}}$ of coherent states of the $\hat{a}_{\bm{q},\sigma}$ operators\cite{Walls_and_Milburn,Serafini}, i.e.~$\hat{a}_{\bm{q}^\prime,\sigma^\prime} \ket{\mathscr{A}}=\alpha_{\bm{q}^\prime,\sigma^\prime}   \ket{\mathscr{A}}$.

The total energy, defined as $E[\{\alpha_{\bm{q},\sigma } \},\psi] \equiv \braket{\Psi|\hat{\cal H}_{\bm B}|\Psi}=\braket{\mathscr{A}| \hat{\cal{H}}^{\rm eff}_{\rm ph}[\psi]|\mathscr{A}}$, is given by:
\begin{eqnarray}\label{eqHtot11}
 &E&[\{\alpha_{\bm{q},\sigma } \},\psi]=\braket{\psi|\hat{\cal H}|\psi}+ \sum_{\bm{q},\sigma } \hbar \omega_{\bm{q}}\left( |\alpha_{\bm{q},\sigma}|^2 + \frac{1}{2} \right)\nonumber +\\&+& \sum_{\bm{q},\sigma } \frac{g\mu_{\rm B}  A_{\bm{q}} }{2} \big[{\bm{S}}  (-\bm{q})  \alpha_{\bm{q},\sigma}  -   {\bm{S}}  (\bm{q})  \alpha_{\bm{q},\sigma}^*        \big]\cdot     iq\bm{u}_{{\rm T},{\bm{q},\sigma}} ~. \nonumber \\
\end{eqnarray} 

Minimization can be performed with respect to $\{\alpha_{\bm{q},\sigma}\}$ analytically by imposing the condition $\partial_{\alpha_{\bm{q},\sigma}^*}E[\{\alpha_{\bm{q},\sigma}\},\psi]=0$. 
We find that the optimal value of $\{\alpha_{\bm{q},\sigma}\}$  is given by:
\begin{eqnarray}
\label{eq:minSpin}
\bar{\alpha}_{\bm{q},\sigma}= \frac{g\mu_{\rm B}  A_{\bm{q}} }{2\hbar \omega_{\rm c} } \braket{\psi| \hat{\bm{S}}({\bm{q} })|\psi}\cdot iq \bm{u}_{{\rm T},{\bm{q},\sigma}}~.
\end{eqnarray}
Note that this equation can be written in terms of the operator
\begin{eqnarray}
\label{eq:EminS9}
\hat{C}_{\bm{q},\sigma} &\equiv &\frac{g\mu_{\rm B}  A_{\bm{q}} }{2\hbar \omega_{\bm{q}} }  \hat{\bm{S}}({\bm{q}}) \cdot iq \bm{u}_{{\rm T},{\bm{q},\sigma}}~,
\end{eqnarray}
i.e.~$\braket{\psi|\hat{C}_{\bm{q},\sigma}|\psi} = \bar{\alpha}_{\bm{q},\sigma} $.
Using Eq.~\eqref{eq:minSpin} into Eq.~(\ref{eqHtot11}) we finally find the energy functional that needs to be minimized with respect to $\ket{\psi}$:
\begin{eqnarray}
\label{eq:minSpin1}
 &E&[\{\bar{\alpha}_{\bm{q},\sigma}\},\psi]=\braket{\psi|\hat{\cal H}|\psi} -\sum_{\bm{q},\sigma} \hbar\omega_{\bm{q}}\Big(|\bar{\alpha}_{\bm{q},\sigma}|^2 - \frac{1}{2} \Big)~.
\end{eqnarray}
Once again, for photon condensation to occur we need $E[\{\bar{\alpha}_{\bm{q},\sigma}\}, \psi]< E[0, \psi_0]$ or, equivalently,
\begin{eqnarray}
\label{eq:EminS1}
\braket{ \psi|\hat{\cal H}|\psi} -\braket{ \psi_0|\hat{\cal H}|\psi_0}< \sum_{\bm{q},\sigma}\hbar\omega_{\bm{q}} |\bar{\alpha}_{\bm{q},\sigma}|^2~.
\end{eqnarray}

As in Sect.~\ref{sect:3D_stiffness_theorem}, the dependence of $\braket{ \psi|\hat{\cal H}|\psi}-\braket{ \psi_0|\hat{\cal H}|\psi_0}$ on $\bar{\alpha}_{\bm{q},\sigma}$ can be calculated exactly up to order $\bar{\alpha}_{\bm{q},\sigma}^2$ by using the stiffness theorem~\cite{Giuliani_and_Vignale}:
\begin{widetext}
\begin{equation}
\label{eq:StiffnessZ}
\braket{ \psi|\hat{\cal H}|\psi}-\braket{ \psi_0|\hat{\cal H}|\psi_0} = -\frac{1}{2}  \sum_{\bm{q},\sigma }  \sum_{\bm{q}^\prime,\sigma^\prime }  \chi^{-1}_{\hat{C}_{\bm{q},\sigma},\hat{C}_{-\bm{q}^\prime,\sigma^\prime}}(0) \bar{\alpha}^*_{\bm{q},\sigma} \bar{\alpha}_{ \bm{q}^\prime,\sigma^\prime}~,
\end{equation}
\end{widetext}
where  $\chi^{-1}_{\hat{C}_{\bm{q},\sigma},\hat{C}_{-\bm{q}^\prime,\sigma^\prime}}(0)$ is the inverse of the static response function $\chi_{\hat{C}_{\bm{q},\sigma},\hat{C}_{-\bm{q}^\prime,\sigma^\prime}}(0)$ and the operator $\hat{C}_{\bm{q},\sigma}$ has been introduced in Eq.~(\ref{eq:EminS9}). Inserting Eq.~(\ref{eq:StiffnessZ}) inside Eq.~(\ref{eq:EminS1}) we find:
\begin{eqnarray}
\label{eq:Stiffness1Z}
- \sum_{\bm{q},\sigma }\left[\frac{1}{ 2\chi_{\hat{C}_{\bm{q},\sigma},\hat{C}_{-\bm{q} ,\sigma}}(0)} +  \hbar  {\omega}_{\bm q}  \right] |\bar{\alpha}_{ \bm{q},\sigma}|^2< 0~.
\end{eqnarray}
Following the same logical steps discussed in Sect.~\ref{sect:3D_stiffness_theorem}, we can consider the previous inequality for a fixed ${\bm q}$:
\begin{eqnarray}
\label{eq:EminS2Z}
 -\chi_{\hat{C}_{\bm{q},\sigma},\hat{C}_{-\bm{q} ,\sigma}} (0)  >\frac{1}{2\hbar  {\omega}_{\bm q} }~.
\end{eqnarray}
We now observe that the homogenous and isotropic nature of the ground state of the 3DES implies~\cite{Giuliani_and_Vignale} $\chi_{\hat{C}_{\bm{q},\sigma},\hat{C}_{-\bm{q}^\prime},\sigma^\prime}(0) =\chi_{\hat{C}_{\bm{q},\sigma},\hat{C}_{-\bm{q},\sigma}}(0)\delta_{\bm{q},\bm{q}^\prime} \delta_{\sigma,\sigma^\prime}$. We readily recognize $\chi_{\hat{C}_{\bm{q},\sigma},\hat{C}_{-\bm{q},\sigma}}(0)$ to be intimately linked to the static, spin-spin response tensor $\chi^{\rm S}_{i,k}({\bm q},0)$.
Indeed, it is easy to show that
\begin{widetext}
\begin{equation}
\label{eq:exact_eigenstate_chiBB1S}
\chi_{\hat{C}_{\bm{q},\sigma},\hat{C}_{-\bm{q},\sigma}}(0)= \frac{q^2g^2\mu^2_{\rm B}  A^2_{\bm{q}} V }{4\hbar^2 \omega^2_{\bm{q}} } \sum_{i,k} {u}^{(i)}_{{\rm T},{\bm{q},\sigma}}  {u}^{(k)}_{{\rm T},{\bm{q},\sigma}} \chi^{\rm S}_{i,k}({\bm q},0)~,
\end{equation}
where
\begin{eqnarray}
\label{eq:exact_chiSS}
\chi^{\rm S}_{i,k}({\bm q},0) =-\frac{1}{V}\sum_{n\neq 0} \frac{\braket{\psi_0|\hat{{S}_i}(-\bm{q})|\psi_n}\braket{\psi_n|\hat{{S}_k}(\bm{q})|\psi_0}}{E_n-E_0}-\frac{1}{V}  \sum_{n\neq 0} \frac{\braket{\psi_0|\hat{{S}_i}({\bm{q}})|\psi_n}\braket{\psi_n|\hat{{S}_k}({-\bm{q}})|\psi_0}}{E_n-E_0} ~,
\end{eqnarray}
and $\hat{{S}_i}({\bm{q}})$, with $i=x,y,z$, denotes the $i$-th Cartesian component of $\hat{\bm{S}}({\bm{q}})$.
\end{widetext}
Isotropy, translational- and spin-rotational invariance imply that the rank-$2$ tensor $\chi^{\rm S}_{i,k}({\bm q},0)$ can be decomposed in terms of the longitudinal, $\chi^{\rm S}_{\rm L}(q,0)$, and transverse, $\chi^{\rm S}_{\rm T}(q,0)$, spin-spin response functions:
\begin{equation}\label{eq:spin-spin-decomposition}
\chi^{\rm S}_{i,k}({\bm q},0) = \chi^{\rm S}_{\rm L}(q,0)\frac{q_i q_k}{q^2}+\chi^{\rm S}_{\rm T}(q,0)\left(\delta_{i,k} -\frac{q_i q_k}{q^2}\right)~.
\end{equation}
Replacing Eq.~(\ref{eq:spin-spin-decomposition}) into Eq.~(\ref{eq:exact_eigenstate_chiBB1S}), we finally find
\begin{equation}\label{eq:chiCC_relation_with_chiST}
\chi_{\hat{C}_{\bm{q},\sigma},\hat{C}_{-\bm{q} ,\sigma}} (0)=\frac{q^2g^2\mu_{\rm B}A^2_{\bm{q}} V }{4\hbar^2 \omega^2_{\bm{q}}  } \chi^{\rm S}_{\rm T}(q,0)~.
\end{equation}

Using Eqs.~\eqref{eq:exact_eigenstate_chiBB1S} and~\eqref{eq:spin-spin-decomposition} and the microscopic expressions of $\omega_{\bm q}=cq/\sqrt{\epsilon_{\rm r}}$ and $A_{\bm q}=\sqrt{2\pi \hbar c^2 /(V \omega_{\bm q} \epsilon_{\rm r})}$ given above, Eq.~\eqref{eq:EminS2Z} can be written as follows:
\begin{eqnarray}
\label{eq:Emin21Z}
-\frac {g^2\mu^2_{\rm B}}{4}~\chi^{\rm S}_{\rm T}(q,0) >\frac{1}{4\pi} ~,
\end{eqnarray}

Again, the left-hand-side of Eq.~\eqref{eq:Emin21Z} has a very clear physical interpretation. It is the non-local transverse spin susceptibility~\cite{Giuliani_and_Vignale}
\begin{eqnarray}\label{eq:nonlocal_spin_susceptibility}
\chi_{\rm spin}(q)\equiv-\frac {g^2\mu^2_{\rm B}}{4}~\chi^{\rm S}_{\rm T}(q,0)~,
\end{eqnarray} 
which, in the long-wavelength $q\to 0$ limit, reduces to the thermodynamic (i.e.~macroscopic) spin magnetic susceptibility (SMS)
\begin{equation}
\chi_{\rm SMS}\equiv \lim_{q\to 0}\chi_{\rm spin}(q)=\left.\frac{\partial M_{\rm S}}{\partial B}\right|_{B=0}~.
\end{equation}
Here, $M_{\rm S}$ is the spin contribution to the magnetization. For free (i.e.~non-interacting) parabolic-band fermions in 3D, $\chi_{\rm SMS}$ reduces to the well-known Pauli spin susceptibility~\cite{Giuliani_and_Vignale}, i.e.
\begin{equation}\label{eq:Pauli_spin_susceptibility}
\chi^{(0)}_{\rm SMS} = \frac{\alpha^2}{r_s }\left(\frac{9}{256\pi^5}\right)^{1/3}>0~,
\end{equation}
where we have used a Land\'e $g$-factor $g=2$. Comparing Eq.~(\ref{eq:Pauli_spin_susceptibility}) with Eq.~(\ref{eq:Landau_diamagnetism}), we find the very well-known result,
\begin{equation}\label{eq:Pauli_in_terms_of_orbital}
\chi^{(0)}_{\rm SMS} = - 3 \chi^{(0)}_{\rm OMS}~.
\end{equation}

In summary, the condition for the occurrence of photon condensation in a 3DES, when the cavity electromagnetic field couples to matter degrees of freedom via the Zeeman coupling only, is:
\begin{eqnarray}\label{eq:EminFinal11}
\boxed{\chi_{\rm spin}(q)>\frac{1}{4\pi}}~.
\end{eqnarray}
\subsection{Combined orbital and Zeeman couplings}
\label{sect:combined_orbital_Zeeman}

In general, when both orbital and spin light-matter interactions are taken into account the total Hamiltonian is:
\begin{eqnarray}\label{eqHtot9Z}
\hat{\cal H}_{{\bm A}+{\bm B}}&=&\hat{\cal H}+ \hat{\cal H}_{\rm ph}+ \frac{g\mu_{\rm B}}{2}\sum_{i=1}^N \hat{\bm{\sigma}}_i\cdot \hat{\bm{B}}(\bm{r}_i)  \nonumber \\
&+&\sum_{i=1}^{N} \frac{e}{ m c}   \hat{{\bm A}}(\bm{r}_i) \cdot \hat{ {\bm p}}_i 
+\sum_{i=1}^{N} \frac{e^2 }{2mc^2}\hat{{\bm A}}^2(\bm{r}_i)~.
\end{eqnarray}
Following the same steps discussed in Sects.~\ref{sect:3D_stiffness_theorem} and~\ref{sect:Zeeman_alone}, one reaches the following condition for the occurrence of photon condensation in a 3DES:
\begin{eqnarray}
\label{eq:EminTot}
-\chi_{\hat{B}_{\bm{q},\sigma}+ \hat{C}_{\bm{q},\sigma},\hat{B}_{-\bm{q},\sigma}+\hat{C}_{-\bm{q},\sigma} }(0)>\frac{1}{2\hbar \Omega_{\bm{q}}}~.
\end{eqnarray}
Now, the key point is that, in the absence of spin-orbit coupling, cross response functions vanish:
\begin{equation}
\chi_{\hat{C}_{\bm{q},\sigma},\hat{B}_{-\bm{q},\sigma} }(0)=\chi_{\hat{B}_{\bm{q},\sigma},\hat{C}_{-\bm{q},\sigma} }(0)=0~.
\end{equation}
This is due to the following facts. Consider for example $\chi_{\hat{C}_{\bm{q},\sigma},\hat{B}_{-\bm{q},\sigma}  }(0)$. We have~\cite{Giuliani_and_Vignale}
\begin{eqnarray}
&&\chi_{\hat{C}_{\bm{q},\sigma},\hat{B}_{-\bm{q},\sigma}  }(\omega) = -\frac{i}{\hbar V} \times \nonumber\\
&\times& \lim_{\eta \to 0} \int_0^\infty d\tau[ \hat{C}_{\bm{q},\sigma}(\tau),\hat{B}_{-\bm{q},\sigma}]e^{i (\omega+i\eta) \tau}~. 
\end{eqnarray}
Since the operators $\hat{C}_{\bm{q},\sigma}(t)$ and $\hat{B}_{-\bm{q},\sigma}$ have disjoint supports, the former acting on the spin degrees of freedom while the latter on the charge degrees of freedom, we have $[\hat{C}_{\bm{q},\sigma}(t),\hat{B}_{-\bm{q},\sigma} ] =0$. We therefore conclude that
\begin{eqnarray}\label{eq:no_cross_terms}
\chi_{\hat{B}_{\bm{q},\sigma}+ \hat{C}_{\bm{q},\sigma},\hat{B}_{-\bm{q},\sigma}+\hat{C}_{-\bm{q},\sigma} }(0) &=& \chi_{\hat{B}_{\bm{q},\sigma},\hat{B}_{-\bm{q},\sigma}}(0) \nonumber\\
&+& \chi_{\hat{C}_{\bm{q},\sigma},\hat{C}_{-\bm{q},\sigma}}(0)~.
\end{eqnarray}
Using Eqs.~(\ref{eq:no_cross_terms}), (\ref{eq:chi_BB_interms_of_transverse_current_response1}), and~(\ref{eq:chiCC_relation_with_chiST}) inside Eq.~(\ref{eq:EminTot}), we find that the condition for occurrence of photon condensation is:
\begin{eqnarray}
\label{eq:EminFinal}
2A^2_{\bm{q}}V\big[ \chi_{\rm orb}(q) +\chi_{\rm spin}(q) \big]q^2>{\hbar \omega_{\bm{q}}}~,
\end{eqnarray}
which, upon substitution of $\omega_{\bm q}=cq/\sqrt{\epsilon_{\rm r}}$ and $A_{\bm q}=\sqrt{2\pi \hbar c^2 /(V \omega_{\bm q} \epsilon_{\rm r})}$, becomes
\begin{eqnarray}
\label{eq:EminFinal2}
\boxed{\chi_{\rm orb}(q) +\chi_{\rm spin}(q) > \frac{1}{4\pi}}~.
\end{eqnarray}
This is the most important result for 3DESs: in the absence of spin-orbit coupling in the matter degrees of freedom---or other microscopic mechanisms that are responsible for non-zero cross response function such as $\chi_{\hat{B}_{\bm{q},\sigma}, \hat{C}_{-\bm{q},\sigma},}(0)$ and $\chi_{\hat{C}_{\bm{q},\sigma},\hat{B}_{-\bm{q},\sigma}  }(0)$---the condition for the occurrence of photon condensation involves the sum of the orbital and spin transverse static response functions.

When electron-electron interactions are negligible (i.e.~$r_{\rm s}\ll 1$), the condition (\ref{eq:EminFinal2}) for the occurrence of 3D photon condensation (i.e.~formation of Condon domains) can be made more explicit. Indeed, consider for example the case of a non-interacting parabolic-band 3D Fermi gas. Using the long-wavelength expression (\ref{eq:Landau_diamagnetism}) and (\ref{eq:Pauli_spin_susceptibility}) inside Eq.~(\ref{eq:EminFinal2}), we immediately see that photon condensation can occur in the absence of electron-electron interactions provided that
\begin{equation}
r_{\rm s} < \left(\frac{2}{3\pi^2}\right)^{1/3}\alpha^2~,
\end{equation}
or, equivalently, provided that the electron density is sufficiently high,
\begin{equation}
n > n_{\rm c} = \frac{9\pi}{8\alpha^6}\frac{1}{a^3_{\rm B}}~.
\end{equation}
Unscreened current-current interactions at low temperatures under strong magnetic fields, which may result in non-Fermi-liquid behavior~\cite{holstein_prb_1973}, lead to the occurrence of long-range magnetic orbital order even at low densities~\cite{gordon_prl_1998}.

\section{2D Photon Condensation}
\label{sect:2D_stiffness_theorem}

In this Section, we consider the problem of a 2DES located in the middle of a quasi-2D cavity.

Similarly to the 3D case discussed above in Sect.~\ref{sect:3D_stiffness_theorem}, 
we describe the 2DES with the jellium model Hamiltonian
\begin{equation}\label{eq:Hamiltonian}
\hat{\cal H}_{\rm 2D}=\sum_{i=1}^N  \frac{\hat{ {\bm p}}_{\parallel,i}^2}{2m} +
\frac{1}{2}\sum_{i\neq j} v(|\hat{\bm r}_{\parallel,i} - \hat{\bm r}_{\parallel,j}|)~,
\end{equation}
where $\hat{\bm r}_{\parallel,i}$ and $\hat{\bm p}_{\parallel,i}$ denote respectively the position and momentum operators of the $i$-th electron moving in the $\hat{\bm x}$-$\hat{\bm y}$ plane. For future use, we introduce the 2D Fourier transforms of the density and paramagnetic (number) current operators:
\begin{eqnarray}
\label{eq:densityq_2D}
\hat{n}(\bm{q}_{\parallel})&=&\sum_{i=1}^N e^{-i\bm{q}_\parallel\cdot \hat{\bm r}_{_\parallel, i}}~,\\
\hat{\bm{j}}_{\rm p}(\bm{q}_{\parallel})&=&\frac{1}{2m} \sum_{i=1}^N \left(\hat{\bm p}_{\parallel, i} e^{-i\bm{q}_{\parallel}\cdot \hat{\bm r}_{\parallel, i}} + e^{-i\bm{q}_{\parallel}\cdot \hat{\bm r}_{\parallel, i}}\hat{\bm p}_{\parallel, i}\right)~,
\end{eqnarray}
with the usual properties $\hat{n}(- \bm{q}_{\parallel})= \hat{n}^\dagger(\bm{q}_{\parallel})$ and $\hat{\bm{j}}_{\rm p}(- \bm{q}_{\parallel})= \hat{\bm{j}}^\dagger_{\rm p}(\bm{q}_{\parallel})$.

We consider a cavity with length $L_{z}$ in the $\hat{\bm z}$ direction, satisfying the quasi-2D condition $L_{z} \ll L_{x},L_{y}$. The walls of the cavity in the $\hat{\bm z}$ direction are assumed to perfectly conducting. Accordingly, the tangential component of the electric field and
the normal component of the magnetic field must vanish at the cavity boundaries~\cite{Kakazu94} $z=\pm L_z/2$. 
In addition, we impose periodic boundary conditions along the $\hat{\bm x}$ and $\hat{\bm y}$  directions. In the Coulomb gauge, the vector potential fulfilling the cavity boundary conditions can be expressed as follows~\cite{Kakazu94}:
\begin{eqnarray}
&\hat{{\bm A}}(\bm{r})&=\sum_{\bm{q}_{\parallel},\sigma, n_z} A^{\rm(2D)}_{\bm{q}_{\parallel} ,n_z} {\bm e}_{\bm{q}_{\parallel},\sigma,n_z } (z) 
 \nonumber \\
&\times&(\hat{a}_{\bm{q}_{\parallel},\sigma ,n_z}e^{i\bm{q}_{\parallel} \cdot \bm{r}_{\parallel} }+\hat{a}_{\bm{q}_{\parallel}, \sigma ,n_z}^\dagger  
e^{-i\bm{q}_{\parallel} \cdot \bm{r}_{\parallel} })~, 
\end{eqnarray}
where
\begin{eqnarray}
{\bm e}_{\bm{q}_{\parallel},1,n_z } (z)&=&{\bm u}_{\bm{q}_{\parallel},1 }  \sin\Big[\frac{\pi n_z}{L_z} \Big(z+\frac{L_z}{2} \Big) \Big]~,\\ 
{\bm e}_{\bm{q}_{\parallel},2,n_z } (z)&=&\frac{\bm q_\parallel}{q_\parallel} \sin\Big[\frac{\pi n_z}{L_z} \Big(z+\frac{L_z}{2} \Big) \Big] 
\frac{\pi n_z}{L_z \sqrt{q_\parallel^2+(\frac{\pi n_z}{L_z})^2}}\nonumber\\
&+& \hat{\bm{z}} 
\cos\Big[\frac{\pi n_z}{L_z} \Big(z+\frac{L_z}{2} \Big)\Big] \frac{i q_\parallel}{ \sqrt{q_\parallel^2+(\frac{\pi n_z}{L_z})^2}}~.\nonumber\\
\end{eqnarray}
Here, $n_z$ is an integer index, ${\bm q}_\parallel= (2\pi n_{x}/L_{x},2\pi n_{y}/L_{y})$ with $(n_{x},n_{y})$ relative integers, $\sigma=1,2$ is the polarization index, 
${\bm u}_{\bm{q}_\parallel,1}$ is the linear polarization vector lying in the $\hat{\bm{x}}$-$\hat{\bm{y}}$ plane and transverse to $\bm {q}_{\parallel}$, i.e.~${\bm u}_{\bm{q}_\parallel,1}\cdot \bm q_{\parallel}=0$,  $A^{\rm(2D)}_{\bm{q}_{\parallel} ,n}=\sqrt{4\pi \hbar c^2 /(L_z S\omega_{\bm{q}_{\parallel} ,n_z} \epsilon_{\rm r})}$,
$S=L_x L_y$,  $\epsilon_{\rm r}$ is the cavity relative dielectric constant,
and $\omega_{\bm{q}_\parallel,n_z}=(c/\epsilon_{\rm r})\sqrt{ q_\parallel^2 + (\pi n_z/L_z)^2}$. In the $\hat{\bm{x}}$-$\hat{\bm{y}}$ plane ($z=0$), where the 2DES lays, modes labeled by the polarization index $\sigma=1$ are transverse waves, i.e.~${\bm e}_{{\bm q}_\parallel,1,n_z}(0)\cdot {\bm q}_\parallel=0$. The second mode labelled by $\sigma = 2$ can be dropped for arbitrarily large wave vector if the 2DES is located exactly in the middle of the cavity since:
\begin{itemize}
\item For odd values of $n_z$, the vector ${\bm e}_{\bm{q}_{\parallel}, 2,~{\rm odd}~n_z} (z=0)$ is longitudinal, i.e.~it is parallel to $\bm{q}_{\parallel}$. 
Since, as a consequence of gauge invariance, the static longitudinal current-current response function is zero~\cite{Giuliani_and_Vignale} for arbitrary $\bm{q}_\parallel$, light-matter interactions with longitudinal photonic modes cannot induce photon condensation. 
\item For even values of $n_z$, the vector ${\bm e}_{\bm{q}_{\parallel},2,~{\rm even}~n_z} (z=0)$ is along the $\hat{\bm{z}}$ direction. Therefore, electronic degrees of freedom cannot couple to modes with $\sigma=2$ and even $n_z$.
\end{itemize}
From now on, we will take into account only modes with $\sigma=1$. In particular, since the 2DES is placed in the middle of the photonic cavity, at $z=0$, only photonic modes with odd $n_z$ couple to the matter degrees of freedom~\cite{Hagenmuller10}. Similarly to the 3D case, the following properties hold true: $\omega_{-{\bm q}_\parallel,n_z}=\omega_{{\bm q}_\parallel,n_z}$, ${\bm u}_{-{\bm q}_\parallel,1}={\bm u}_{{\bm q}_\parallel,1}$, $A^{\rm(2D)}_{-{\bm q}_\parallel,n_z}=A^{\rm(2D)}_{{\bm q}_\parallel,n_z}$.

The Hamiltonian of the 2DES coupled to the cavity field is expressed as
\begin{eqnarray}\label{eqHtot2d}
&\hat{\cal H}_{{\bm A}}&=\hat{\cal H}_{\rm 2D}+\hat{\cal H}_{\rm ph}
+\sum_{i=1}^{N} \frac{e}{ m c}   \hat{{\bm A}}(\bm{r}_{{\parallel},i},z=0) \cdot \hat{ {\bm p}}_{{\parallel},i}\nonumber \\
&+&\sum_{i=1}^{N} \frac{e^2 }{2mc^2}\hat{\bm A}^2(\bm{r}_{{\parallel},i},z=0)~,
\end{eqnarray}
where the cavity Hamiltonian $\hat{\cal H}_{\rm ph}$ reads as following
\begin{equation}\label{eq:photon_part_Hamiltonian_2D}
\hat{\cal H}_{\rm ph}= \sum_{\bm{q}_\parallel, \sigma, n_z}  \hbar\omega_{\bm{q}_{\parallel},n_z}\hat{a}_{\bm{q}_{\parallel},\sigma,n_z}^\dagger\hat{a}_{\bm{q}_{\parallel},\sigma,n_z}~.
\end{equation}
This needs to be compared with the 3D one in Eq.~(\ref{eqHtot}). Once again, the third and the fourth term in Eq.~({\ref{eqHtot2d}}) are the paramagnetic and diamagnetic contributions, respectively. A constant term in Eq.~(\ref{eq:photon_part_Hamiltonian_2D}) has been dropped, since below we will be only interested in energy differences. 
From now on, we will follows steps similar to those described in Sect.~\ref{sect:3D_stiffness_theorem}. We will therefore mainly highlight differences between the 3D case discussed there and the 2D case discussed in this Section and cut short on the algebraic steps that are identical in the two cases. On purpose, and with notational abuse, we will denote by the same symbols  quantities that in both cases have an identical physical meaning.

As in the 3D case, we are interested in the possible occurrence of a quantum phase transition to a photon condensate, and we therefore wish to make general 
statements about  the ground state $\ket{\Psi}$ of $\hat{\cal H}_{{\bm A}}$,  in the 2D thermodynamic limit $N\to\infty$, $S\to \infty$, with constant $n_{\rm 2D} = N/S$. 
In this limit, we can safely assume that $\ket{\Psi}$ does not contain light-matter entanglement, 
i.e.~we can take $\ket{\Psi}=\ket{\psi}\ket{\Phi}$, where $\ket{\psi}$ and $\ket{\Phi}$ are matter and light states.  The effective Hamiltonian for the photonic degrees of freedom is 
$\hat{\cal{H}}^{\rm eff}_{\rm ph}[{\psi}]\equiv   \braket{\psi| \hat{\cal H}_{{\bm A}} |\psi}$. 
The order parameter for 2D photon condensation is $\bar{\alpha}_{{\bm q}_\parallel,1,n_z}\equiv\braket{\Phi|\hat{a}_{{\bm q}_\parallel,1,n_z}|\Phi}$, which, at the putative QCP, is small. 
Since the diamagnetic term in Eq.~(\ref{eqHtot2d}) is quadratic in $\bar{\alpha}_{{\bm q}_\parallel,1,n_z}$, 
close to the QCP we can approximate the matter content in the diamagnetic term with its value in the absence of light-matter interactions. By further assuming, as in the 3D case, that the ground state of the 2DES in the absence of light-matter interactions is homogenous and isotropic, i.e.~that $\braket{\psi_0|\hat{n}({\bm q}_{\parallel})|\psi_0}=N\delta_{{\bm q}_{\parallel}, {\bm 0}}$,
the effective photon Hamiltonian can be written as
\begin{eqnarray}\label{eqHph2d}
&\hat{\cal{H}}^{\rm eff}_{\rm ph}&[{\psi}]=\braket{ \psi|\hat{\cal H}_{\rm 2D}|\psi} +
\hat{\cal H}_{\rm ph}+\hat{\cal H}_{\rm p}+\hat{\cal H}_{\rm d}~,
\end{eqnarray}
where the paramagnetic contribution is given by
\begin{eqnarray}\label{eq:paramagnetic_2D}
&\hat{\cal H}_{\rm p}& =\sum_{{\rm odd}~n_z} \sum_{{\bm q}_\parallel}
(-1)^{\frac{n_z-1}{2}}\frac{e}{c}A^{({\rm 2D})}_{\bm{q}_{\parallel}, n_{z}}
\Big[ \hat{a}_{\bm{q}_\parallel,1,n_z} {\bm u}_{{\bm q}_{\parallel},1}\cdot {\bm j}_{\rm p}(-{\bm q}_\parallel)  \nonumber \\
&+&\hat{a}^\dagger_{\bm{q}_\parallel,1, n_z} {\bm u}_{{\bm q}_{\parallel},1}\cdot {\bm j}_{\rm p}({\bm q}_\parallel)\Big] 
\end{eqnarray}
and the diamagnetic one by
\begin{eqnarray}\label{eq:diamagnetic_2D}
&\hat{\cal H}_{\rm d}& = \sum_{{\rm odd}~n_z,n_z^\prime} \sum_{{\bm q}_\parallel}(-1)^{\frac{n_z+n_z^\prime - 2}{2}}\frac{e^2}{2 m c^2}A^{({\rm 2D})}_{\bm{q}_{\parallel}, n_{z}}A^{({\rm 2D})}_{\bm{q}_{\parallel}, n^\prime_{z}}\nonumber\\
&\times&(\hat{a}_{\bm{q}_\parallel,1, n_z}+\hat{a}^\dagger_{-\bm{q}_\parallel,1, n_z})(\hat{a}^\dagger_{\bm{q}_\parallel, 1, n_z^\prime}+\hat{a}_{-\bm{q}_\parallel, 1, n_z^\prime})~.
\end{eqnarray}
In Eq.~(\ref{eq:paramagnetic_2D}) we have introduced
\begin{equation}
{\bm j}_{\rm p}({\bm q}_\parallel) \equiv \braket{\psi|\hat{\bm j}_{\rm p}({\bm q}_\parallel) |\psi}~.
\end{equation}
For future use, we also introduce ${\cal J}_{{\bm q}_\parallel,1}={\bm u}_{{\bm q}_{\parallel},1} \cdot {\bm j}_{\rm p}({\bm q}_\parallel)$.

As we have seen in Sect.~\ref{sect:3D_discussion}, point v), in order to calculate the energy functional, it is sufficient to evaluate the expectation value of the effective Hamiltonian $\hat{\cal{H}}^{\rm eff}_{\rm ph}[{\psi}]$ on a trial photonic wavefunction of the form $\ket{\mathscr{A}} \equiv \otimes_{\bm{q},n_z} \ket{\alpha_{\bm{q},1,n_z}}$, namely on a tensor product of coherent states of the $\hat{a}_{\bm{q},1,n_z}$ operators, i.e.~$\hat{a}_{\bm{q}^\prime,1,n_z} \ket{\mathscr{A}}=\alpha_{\bm{q}^\prime,1,n_z}\ket{\mathscr{A}}$. This procedure corresponds to replacing the photonic operators in Eq.~\eqref{eqHLAbb2}  with $c$-numbers, $\hat{a}_{{\bm q}_\parallel,1,n_z} \to  \alpha_{{\bm q}_\parallel,1,n_z}$.
Carrying out this procedure we find:
\begin{widetext}
\begin{eqnarray}
\label{eq:Ealpha2D}
&E&[\{ {\alpha}_{{\bm q}_\parallel,1,n_z} \}, \psi]=\braket{\psi|\hat{\cal H}_{\rm 2D} |\psi} +\sum_{{\rm odd}~n_z} \sum_{{\bm q}_\parallel} \frac{(-1)^{\frac{n_z-1}{2}} \sqrt{2D}}{\sqrt{\omega_{{\bm q}_\parallel, n_z}}} 
\Big[\alpha_{\bm{q}_\parallel,1,n_z} {\cal J}_{-{\bm q}_\parallel,1} +{\rm c.c.} \Big] \nonumber \\ 
&+& \sum_{{\rm odd}~n_z}  \sum_{{\rm odd}~n^\prime_z} \sum_{{\bm q}_\parallel}  \frac{(-1)^{\frac{n_z+n_z^\prime-2}{2}} D N}{m \sqrt{\omega_{{\bm q}_\parallel, n_z} \omega_{{\bm q}_\parallel, n_z^\prime} } } (\alpha_{\bm{q}_\parallel, 1, n_z}+\alpha^*_{-\bm{q}_\parallel, 1, n_z})(\alpha^*_{\bm{q}_\parallel, 1, n_z^\prime}+\alpha_{-\bm{q}_\parallel, 1, n_z^\prime})+ \sum_{\bm{q}_\parallel,1,n_z}  \hbar\omega_{\bm{q}_{\parallel},n_z}\alpha_{\bm{q}_{\parallel},1, n_z}^*\alpha_{\bm{q}_{\parallel},1, n_z}~, \nonumber \\
\end{eqnarray}
\end{widetext}
where $D\equiv 2 \pi \hbar e^2/(L_z S \epsilon_{\rm r})$.
(As discussed in Sect.~\ref{sect:3D_stiffness_theorem}, if one is interested in finding the exact photonic eigenstate, a different and much more cumbersome root needs to be followed. This is described at length in Appendix \ref{app:details_bog} and related Appendix~\ref{app:calculation_determinant}. The end result, from the point of view of energy differences, is identical to the one that one obtains using Eq.~\eqref{eq:Ealpha2D}.) Note that all the modes with even $n_z$ are completely decoupled from matter degrees of freedom. For these modes, the minimum of the energy functional is trivially obtained at $\alpha_{\bm{q}_\parallel, 1, n_z}=0$. Hence, we can completely disregard even values of $n_z$: from now on, the index $n_z$ will take only odd values. 

It turns out to be useful to express the energy functional $E[\{\alpha_{\bm{q}_\parallel, 1,n_z}\}, \psi]$ in terms of $\{{\bm z}_{{\bm q}_\parallel,1,n_z}\}=\{( x_{{\bm q}_\parallel,1, n_z}, y_{{\bm q}_\parallel,1,n_z})^\top\}$
where $x_{{\bm q}_\parallel,1, n_z}=(\alpha_{{\bm q}_\parallel,1,n_z}+ \alpha^\ast_{-{\bm q}_\parallel,1,n_z})/2$ and $y_{{\bm q}_\parallel,1, n_z}=(\alpha_{{\bm q}_\parallel,1,n_z}- \alpha^\ast_{-{\bm q}_\parallel,1,n_z})/(2i)$. Introducing $g_j(\bm{q}_\parallel)=(-1)^{j}\sqrt{2D/\omega_{\bm{q}_\parallel,2j+1}}$, we find 
\begin{eqnarray}
&&E[\{ {\bm z}_{{\bm q}_\parallel,1,n_z} \}, \psi]=
\braket{\psi|\hat{\cal H}_{\rm 2D} |\psi}+ \sum_{{\bm q}_\parallel,1} \sum_{{\rm odd}~ n_z}  \Big[  \hbar \omega_{{\bm q}_\parallel,n_z}  \nonumber \\
&\times&(x_{{\bm q}_\parallel,1,n_z}x_{-{\bm q}_\parallel,1,n_z}+y_{{\bm q}_\parallel,1,n_z}y_{-{\bm q}_\parallel,1,n_z}) \nonumber \\
&+& \frac{2N}{m} \sum_{{\rm odd}~n_z^\prime} g_{(n_z-1)/2}({\bm q}_\parallel) g_{(n_z^\prime-1)/2}({\bm q}_\parallel) x_{{\bm q}_\parallel,1,n_z}x_{-{\bm q}_\parallel,1,n_z^\prime}
\nonumber\\
&+& 2{\cal J}_{-{\bm q}_\parallel,1}   g_{(n_z-1)/2}({\bm q}_\parallel)  x_{{\bm q}_\parallel,1,n_z}\Big]~.
\end{eqnarray}
This needs to be minimized with respect to $\{ {\bm z}_{{\bm q}_\parallel,1,n_z} \}$ and $|\psi\rangle$.  
The minimization with respect to $\{ {\bm z}_{{\bm q}_\parallel,1,n_z} \}$  can be done analytically by imposing the condition 
$\partial_{  {\bm z}_{{\bm q}_\parallel,1,n_z} ^*}E[\{ {\bm z}_{{\bm q}_\parallel,1,n_z}\}, \psi]=0$. 
We find that the optimal value of  $\{  {\bm z}_{{\bm q}_\parallel,1,n_z} \}$ is given by: $\hbar \omega_{{\bm q}_\parallel,n_z}   y_{{\bm q}_\parallel,1,n_z}=0$ and 
\begin{widetext}
\begin{equation}\label{eq:SRx}
\hbar \omega_{{\bm q}_\parallel,n_z} x_{{\bm q}_\parallel,1,n_z}+\frac{2N}{m}\sum_{{\rm odd}~n_z^\prime} g_{(n_z-1)/2}({\bm q}_\parallel)    
 g_{(n_z^\prime-1)/2}({\bm q}_\parallel)  x_{{\bm q}_\parallel,1,n_z^\prime}=
-g_{(n_z-1)/2}({\bm q}_\parallel)  {\cal J}_{{\bm q}_\parallel,1}~,
\end{equation}
\end{widetext}
where $n_z$ is odd.

The first equation is trivially solved by $y_{{\bm q}_\parallel,1,n_z}=0$. From Eq.~(\ref{eq:SRx}), we find that the optimal value of  $\{x_{{\bm q}_\parallel,1,n_z} \}$ is 
the solution of a linear system in terms of $ {\cal J}_{{\bm q}_\parallel,1} $, and
it is non-trivial (i.e.~$x_{{\bm q}_\parallel,1,n_z}\neq 0$) only if ${\cal J}_{{\bm q}_\parallel,1}$ takes a finite value.
Using the stiffness theorem~\cite{Giuliani_and_Vignale}, one has, up to second order in ${\cal J}_{{\bm q}_\parallel,1}$,
\begin{widetext}
\begin{eqnarray}
 \braket{\psi| \hat{\cal H}_{\rm 2D} |\psi} -  \braket{\psi_0| \hat{\cal H}_{\rm 2D} |\psi_0} = -  \frac{1}{2 S} \sum_{{\bm q}_\parallel,{\bm q}^\prime_\parallel} 
 \chi^{-1}_{{\bm u}_{{\bm q}_\parallel,1}\cdot \hat{\bm j}_{\rm p}(\bm{q}_\parallel) , {\bm u}_{{\bm q}_\parallel^\prime,1} \cdot \hat{\bm j}_{\rm p}(-\bm{q}^\prime_\parallel)}(0)  
 {\cal J}_{{\bm q}_\parallel,1}  {\cal J}_{-{\bm q}_\parallel^\prime,1}~.
\end{eqnarray}
In writing the previous equation we have assumed, as in the 3D case, that $\braket{\psi_0|\hat{\bm{j}}_{\rm p}(\bm{q}_\parallel)|\psi_0}=0$ for all values of $\bm{q}_\parallel$. 
Since the ground state of the 2DES has been taken to be homogenous and isotropic, the following property holds true:
\begin{equation}\label{eq:chi_jj_interms_of_paramagnetic_tensor}
\chi_{{\bm u}_{{\bm q}_\parallel,1} \cdot \hat{\bm j}_{\rm p}(\bm{q}_\parallel), {\bm u}_{{\bm q}_\parallel^\prime,1} \cdot \hat{\bm j}_{\rm p}(-\bm{q}^\prime_\parallel)}(0) 
= \chi_{{\bm u}_{{\bm q}_\parallel,1} \cdot \hat{\bm j}_{\rm p}(\bm{q}_\parallel) , {\bm u}_{{\bm q}_\parallel,1} \cdot \hat{\bm j}_{\rm p}(-\bm{q})}(0) 
 \delta_{{\bm q}_\parallel, {\bm q}_\parallel^\prime }~.
\end{equation}
\end{widetext}
Similarly to the 3D case, we now express the response function $\chi_{{\bm u}_{{\bm q}_\parallel,1} \cdot \hat{\bm j}_{\rm p}(\bm{q}_\parallel) , {\bm u}_{{\bm q}_\parallel,1} \cdot \hat{\bm j}_{\rm p}(-\bm{q})}(0)$ in terms of the {\it  physical}~ current-current response tensor~\cite{Giuliani_and_Vignale}, 
which contains a paramagnetic as well as a diamagnetic contribution:
\begin{eqnarray}\label{eq:chi_phys2d}
\chi^{\rm J}_{i,k}(\bm{q}_\parallel,0)=\frac{n_{\rm 2D}}{m}\delta_{i,k} +\chi_{\hat{j}_{{\rm p}, i}(\bm{q}),\hat{j}_{{\rm p}, k}(-\bm{q}_\parallel)}(0)~. \nonumber \\
\end{eqnarray}
Since we are considering  a homogeneous and isotropic system, the rank-$2$ tensor $\chi^{\rm J}_{i,k} (\bm{q}_\parallel,0)$ 
can be decomposed in terms of the longitudinal, $\chi^{\rm J}_{\rm L}(q_\parallel,0)$, and transverse, $\chi^{\rm J}_{\rm T}(q_\parallel,0)$, current-current response functions~\cite{Giuliani_and_Vignale}:
\begin{eqnarray}\label{eq:decomposition long and trans 2D}
\chi^{\rm J}_{i,k}(\bm{q}_\parallel,0) &=& \chi^{\rm J}_{\rm L}(q_\parallel,0)\frac{q_{\parallel, i} q_{\parallel, k}}{q_\parallel^2} \nonumber\\
&+& \chi^{\rm J}_{\rm T}(q_\parallel,0)\left(\delta_{i,k} -\frac{q_{\parallel, i} q_{\parallel, k}}{q_\parallel^2}\right)~.
\end{eqnarray}
Note that, as a consequence of gauge invariance, $\chi^{\rm J}_{\rm L}(q_\parallel,0)=0$ for every $q_\parallel$~\cite{Giuliani_and_Vignale}.
Using Eqs.~(\ref{eq:chi_phys2d})-(\ref{eq:decomposition long and trans 2D}) in Eq.~(\ref{eq:chi_jj_interms_of_paramagnetic_tensor}), we finally find
\begin{equation}\label{eq:chi_BB_interms_of_transverse_current_response}
\chi_{{\bm u}_{{\bm q}_\parallel,1} \cdot \hat{\bm j}_{\rm p}(\bm{q}_\parallel) , {\bm u}_{{\bm q}_\parallel,1} \cdot \hat{\bm j}_{\rm p}(-\bm{q})}(0) = 
\left[\chi^{\rm J}_{\rm T}(q_\parallel,0)-\frac{n_{\rm 2D}}{m}\right]~.
\end{equation}

We now calculate the energy difference between a generic phase with $[ {\bm z}_{{\bm q}_\parallel,1} , \psi]$ and the normal phase with
$[ {\bm z}_{{\bm q}_\parallel,1}={\bm 0} , \psi_0]$ (where $ {\bm z}_{{\bm q}_\parallel,1}=\{ {\bm z}_{{\bm q}_\parallel,1,n_z}\}_{{\rm odd}~n_z}$):
\begin{eqnarray}\label{eq:SR-N}
&&
E[ {\bm z}_{{\bm q}_\parallel,1} , \psi]-E[ {\bm z}_{{\bm q}_\parallel,1}={\bm 0} , \psi_0]= 
\nonumber\\
&& \sum_{{\bm q}_\parallel} \Big\{
\frac{1}{2S}\left[\frac{n_{\rm 2D}}{m}-\chi^{\rm J}_{\rm T}(q_\parallel,0)\right]^{-1}  {\cal J}_{{\bm q}_\parallel,1}  {\cal J}_{-{\bm q}_\parallel,1}\nonumber \\
&+& \sum_{{\rm odd}~ n_z}  \Big[  \hbar \omega_{{\bm q},n_z}
(x_{{\bm q}_\parallel,1,n_z}x_{-{\bm q}_\parallel,1,n_z}+y_{{\bm q}_\parallel,1,n_z}y_{-{\bm q}_\parallel,1,n_z}) \nonumber \\
&+& \frac{2N}{m} \sum_{{\rm odd}~n_z^\prime} g_{(n_z-1)/2}({\bm q}_\parallel) g_{(n_z^\prime-1)/2}({\bm q}_\parallel) x_{{\bm q}_\parallel,1,n_z}x_{-{\bm q}_\parallel,1,n_z^\prime}
\nonumber\\
&+& 2{\cal J}_{-{\bm q}_\parallel,1}   g_{(n_z-1)/2}({\bm q}_\parallel)  x_{{\bm q}_\parallel,1,n_z}\Big]\Big\}~.
\end{eqnarray}
Minimizing this quantity with respect to ${{\cal J}_{{\bm q}_\parallel,1} }$, we obtain the following result:
\begin{eqnarray}\label{eq:minimum_constraint_2D}
{\cal J}_{{\bm q}_\parallel,1} &=& 2 S
\left[\chi^{\rm J}_{\rm T}(q_\parallel,0)-\frac{n_{\rm 2D}}{m}\right] \times \nonumber \\
&\times& \sum_{{\rm odd}~n_z}  g_{(n_z-1)/2}({\bm q}_\parallel)  x_{{\bm q}_\parallel,1,n_z}~.
\end{eqnarray}

Replacing Eq.~(\ref{eq:minimum_constraint_2D}) in Eq.~(\ref{eq:SR-N}), we find that the energy difference, minimized with respect to the matter wave-function and denoted by $ {\cal E}[  {\bm z}_{{\bm q}_\parallel,1} ]\equiv \min_\psi \big( E[ {\bm z}_{{\bm q}_\parallel,1} , \psi]-E[ {\bm z}_{{\bm q}_\parallel,1}={\bm 0} , \psi_0]  \big)$ takes the following quadratic form:
\begin{eqnarray}\label{eq:SR-N-quadratic}
&& {\cal E}[ {\bm z}_{{\bm q}_\parallel,1} ]=  \sum_{{\bm q}_\parallel} \sum_{{\rm odd}~ n_z}  
\Big[ \hbar \omega_{{\bm q_\parallel}, n_z}  (x_{{\bm q}_\parallel,1,n_z}x_{-{\bm q}_\parallel,1,n_z}\nonumber \\
&+&  y_{{\bm q}_\parallel,1,n_z}y_{-{\bm q}_\parallel,1,n_z}) +  \frac{2S}{m} \chi^{\rm J}_{\rm T}(q_\parallel,0)  \sum_{{\rm odd}~n_z^\prime} g_{(n_z-1)/2}({\bm q}_\parallel)\nonumber \\
&\times& g_{(n_z^\prime-1)/2}({\bm q}_\parallel) x_{{\bm q}_\parallel,1,n_z}x_{-{\bm q}_\parallel,1,n_z^\prime}
\Big]~,
\end{eqnarray}
which can be written compactly as 
\begin{equation}\label{eq:calM}
{\cal E}[ {\bm z}_{{\bm q}_\parallel,1} ]=  \sum_{{\bm q}_\parallel}  
{\bm z}_{{\bm q}_\parallel,1}^\dagger {\cal M}_{{\bm q}_\parallel} {\bm z}_{{\bm q}_\parallel,1}~.
\end{equation}
Here, ${\cal M}_{{\bm q}_\parallel}$ is a symmetric matrix.
For photon condensation to occur we need the photon condensate phase to be energetically favored with respect to the normal phase. This occurs, at a given ${\bm q}_{\parallel}$, if at least one eigenvalue $\lambda_{{\bm q}_\parallel, n}$ of ${\cal M}_{{\bm q}_\parallel}$ is negative. For each ${\bm q}_\parallel$, the determinant $\Delta_{{\bm q}_\parallel}={\rm Det}({\cal M}_{{\bm q}_\parallel} )$ of the quadratic form in Eq.~(\ref{eq:calM}) can be written as (see Appendix~\ref{app:determinant_2d}):
\begin{eqnarray}\label{eq:determinant_2D_stiffness}
\Delta_{{\bm q}_\parallel} &=& \left[  1+ \chi^{\rm J}_{\rm T}(q_\parallel,0) \frac{2 \pi e^2}{c^2 q_\parallel} 
\tanh\Big(\frac{q_\parallel L_z}{2}\Big)  \right] \times \nonumber\\
&\times&\prod_{{\rm odd}~n_z} (\hbar  \omega_{{\bm q}_\parallel, n_z})^2~.
\end{eqnarray}
Using the relation $\Delta_{{\bm q}_\parallel} = \prod_n \lambda_{{\bm q}_\parallel, n}$ between eigenvalues and determinant, and noting that the second line in Eq.~(\ref{eq:determinant_2D_stiffness}) is positive definite, we conclude that, in order to have at least one negative eigenvalue, the following inequality needs to be satisfied:
\begin{equation}\label{eq:2d_condition}
- \chi^{\rm J}_{\rm T}(q_\parallel,0) \frac{2 \pi e^2}{c^2 q_\parallel} 
\tanh\Big(\frac{q_\parallel L_z}{2}\Big) > 1~.
\end{equation}
This equation generalizes the criterion for photon condensation obtained in Ref.~\onlinecite{Basko19} for the case of a 2DES with Rashba spin-orbit coupling, placed in an external uniform magnetic field.

Let us consider first the case of zero photon momentum, $q_\parallel = 0$. In this case, the condition (\ref{eq:2d_condition}) for the occurrence of the photon condensation reduces to 
\begin{equation}\label{eq:2d_nogo}
- \chi^{\rm J}_{\rm T}(0,0) \frac{ \pi e^2 L_z}{c^2} > 1~.
\end{equation}
As discussed in Sect.~\ref{sect:3D_stiffness_theorem}, in systems with no long-range order~\cite{Giuliani_and_Vignale}, $\lim_{q_\parallel \to 0} \chi^{\rm J}_{\rm T}(q_\parallel,0)=0$. Such diamagnetic sum-rule then yields an absurd ($0>1$), expressing the no-go theorem for the occurrence of  
photon condensation in a spatially-uniform cavity field. 

As in the 3D case, we now introduce the 2D non-local orbital susceptibility
\begin{equation}
\chi_{\rm orb}(q_\parallel) \equiv -\frac{e^2}{c^2}\frac{\chi^{\rm J}_{\rm T}(q_{\parallel},0)}{q^2_{\parallel}}~.
\end{equation}
Introducing this definition in Eq.~(\ref{eq:2d_condition}), we finally obtain the condition for the occurrence of photon condensation in a 2DES:
\begin{equation}\label{eq:2d_condition1}
\boxed{\chi_{\rm orb}(q_\parallel) > \frac{1}{2 \pi q_{\parallel}
\tanh(q_\parallel L_z/2)}}~.
\end{equation}
This is the most important result of this Section. 

As in the 3D case discussed in Sect.~\ref{sect:3D_stiffness_theorem}, the criterion in Eq.~(\ref{eq:2d_condition1}) emphasizes that the route towards the discovery of photon condensate states relies entirely on the knowledge of the orbital magnetic response function $\chi_{\rm orb}$ of ESs.

\subsection{Discussion}
\label{sect:2D_discussion}

In order to gain a deeper understanding on the possible occurrence of 2D photon condensation, we multiply both sides of Eq.~\eqref{eq:2d_condition1} by $2/L_z$ and re-write the criterion as following:
\begin{equation}\label{eq:2d_condition1_dimensionless}
\frac{2\chi_{\rm orb}(q_\parallel)}{L_z} > \frac{1}{2 \pi (q_{\parallel}L_z/2)
\tanh(q_\parallel L_z/2)}~.
\end{equation}
Note that, in this form, both sides of the inequality are dimensionless. We now discuss two regimes of $q_\parallel$ (short-wavelength and long-wavelength regimes) where Eq.~(\ref{eq:2d_condition1_dimensionless}) can be satisfied. 

The right-hand side of Eq.~(\ref{eq:2d_condition1_dimensionless}) decreases with increasing $q_{\parallel}$. It therefore seems easy to satisfy Eq.~(\ref{eq:2d_condition1_dimensionless}) at short wavelengths, i.e.~at $q_{\parallel} =1/\ell_{\rm matter}$, where $\ell_{\rm matter}$ is a characteristic microscopic length scale of the 2DES at hand~\cite{footnote_l_matter}. Indeed, since $\ell_{\rm matter}$ is expected to be $\ll L_z/2$, the right-hand side of Eq.~(\ref{eq:2d_condition1_dimensionless}) is small at $q_\parallel \sim 1/\ell_{\rm matter}$ and the threshold condition for 2D photon condensation reduces to
\begin{equation}
\frac{2\pi \chi_{\rm orb}(q_\parallel = 1/\ell_{\rm matter})}{\ell_{\rm matter}} \gtrsim 1~,
\end{equation}
where we have used that $\tanh(L_z/(2\ell_{\rm matter}))\sim 1$.
It may be however very  inconvenient to hunt for 2D photon condensation at wave number scales on the order of $1/\ell_{\rm matter}$, as this would require cavities operating at very high energies, on the order of $\hbar \omega \sim \hbar c q_\parallel/\epsilon_{\rm r}  = \hbar c/(\epsilon_{\rm r} \ell_{\rm matter})$.

From the argument above, it is advisable to investigate whether the 2D criterion (\ref{eq:2d_condition1_dimensionless}) can be satisfied in the long-wavelength $q_\parallel \to 0$ limit. In this respect, we invite the reader to compare Eq.~(\ref{eq:2d_condition1_dimensionless}) with the 3D criterion in Eq.~(\ref{eq:Stiffness7}). The two criteria display a dramatic qualitative difference. While in the 3D case photon condensation can occur also in the quasi-homogeneous $q\to 0$ limit (provided that Eq.~(\ref{eq:Stiffness7}) is satisfied in that limit), in the 2D case the right-hand side of Eq.~\eqref{eq:2d_condition1_dimensionless} diverges as $1/q_{\parallel}^2$ in the $q_{\parallel} \to 0$ limit. On the other hand, the left-hand side is usually finite in the same limit. At a first, superficial glance, it therefore seems impossible to satisfy the condition (\ref{eq:2d_condition1_dimensionless}) in the long-wavelength limit.

However, a useful intermediate small-$q_\parallel$ regime exists. Indeed, the quantity $\chi_{\rm orb}(q_\parallel)$ on the left-hand side of Eq.~\eqref{eq:2d_condition1_dimensionless} is expected to change on a wave number scale controlled by $1/\ell_{\rm matter}$. Matter is in the quasi-homogenous $q_{\parallel} \to 0$ limit when $q_{\parallel}\ll 1/\ell_{\rm matter}$. On the other hand, the right-hand side of Eq.~(\ref{eq:2d_condition1_dimensionless}) changes when $q_{\parallel}$ changes relatively to $2/L_{z}$. In order to mitigate the growth of the right-hand side of Eq.~(\ref{eq:2d_condition1_dimensionless}) with decreasing $q_{\parallel}$, it is therefore wise to work in the regime
\begin{equation}\label{eq:quasihom_limit}
\frac{2}{L_z} \lesssim q_{\parallel} \ll \frac{1}{\ell_{\rm matter}}~, 
\end{equation}
assuming, as above, that $L_z/2 \gg \ell_{\rm matter}$.

When $q_{\parallel}\sim 2/L_{z}\ll 1/\ell_{\rm matter}$, the right-hand side of Eq.~(\ref{eq:2d_condition1_dimensionless}) is $\approx [2\pi \tanh(1)]^{-1}$, and the criterion for 2D photon condensation reduces to
\begin{equation}\label{eq:2d_condition1_dimensionless_practical}
\frac{2\chi_{\rm OMS}}{L_z} \gtrsim 0.21~,
\end{equation}
where, in analogy to the 3D case in Eq.~(\ref{eq:OMS_3D}),
\begin{equation}
\chi_{\rm OMS}\equiv \lim_{q_\parallel \to 0}\chi_{\rm orb}(q_\parallel)~.
\end{equation}
In summary, in order to satisfy the inequality (\ref{eq:2d_condition1_dimensionless}) in the quasi-homogeneous regime (\ref{eq:quasihom_limit}), we need to hunt for 2DESs whose OMS is positive (orbital paramagnets) and larger than $\approx L_{z}/10$.

We now list 2DESs where the criterion (\ref{eq:2d_condition1_dimensionless_practical}) is most likely to be satisfied. In 1991, Vignale demonstrated~\cite{Vignale91} that when the Fermi energy is sufficiently close to a saddle point of the band structure, non-interacting 2DESs in a periodic potential display orbital paramagnetism with $\chi_{\rm OMS}$ diverging logarithmically. The divergence is due to a diverging density of states at the saddle point. The positive sign of $\chi_{\rm OMS}$ is an exquisite quantum effect, which is easy to understand. Near a saddle point the semiclassical approximation breaks down, and tunnelling from one quasi-classical trajectory to the neighboring one occurs. Due to tunneling, electrons rotate around the saddle point in a direction opposite to the classical direction of rotation and the induced magnetic moment is reversed.  We emphasize that the positive sign (i.e.~paramagnetic character of the response) for non-interacting electrons is surprising, in view of the fact that non-interacting {\it parabolic-band} ESs are characterized by a negative OMS (Landau diamagnetism). Recently discovered~\cite{Fu} high-order van Hove singularities are expected to give stronger-than-logarithmic orbital paramagnetic behavior.

More recently, the OMS of the 2DES in graphene has received some attention. In the massless Dirac fermion continuum model, the 2DES in graphene is strongly diamagnetic~\cite{mcclure_pr_1956}, $\chi_{\rm OMS} \propto -\delta(E_{\rm F})$, when the Fermi energy lies at the Dirac point and electron-electron interactions are neglected. On the other hand, the lattice contribution~\cite{stauber_prl_2011} to the OMS beyond the massless Dirac fermion continuum model is positive for a wide range of Fermi energies and diverges at the saddle point, in agreement with Ref.~\onlinecite{Vignale91}. Electron-electron interactions display the same tendency and, in the massless Dirac fermion continuum model, turn the 2DES in graphene into an orbital paramagnet~\cite{Principi10} when the Fermi energy is away from the Dirac point.

The OMS of multi-band systems with a pair of Dirac points interpolating between honeycomb and dice lattices has been studied by Raoux et al.~\cite{Raoux14}. Orbital paramagnetic behavior, stemming from a topological Berry phase changing continuously from $\pi$ (graphene) to $0$ (dice), has been found in this work even at Dirac crossings. A novel geometric contribution to the OMS has been shown to give rise to very strong orbital paramagnetism in models with flat bands~\cite{Raoux16}. It is therefore very natural to expect the same behavior also in twisted bilayer graphene close to the magic angle~\cite{bistritzer_pnas_2011}. 

Other instances of orbital paramagnetic behavior have been found recently in a non-interacting 2DES in the presence of Rashba spin-orbit coupling and a perpendicular static magnetic field~\cite{Basko19}. In particular, in their model, Nataf et al.~\cite{Basko19} showed that Eq.~(\ref{eq:2d_condition1}) is satisfied at $q_{\parallel} \sim 1/\ell_{\rm B}$, every time that two Landau levels with opposite helicity cross.

\section{Summary and conclusions}
\label{sect:summary_and_conclusions}

In summary, we have derived criterions for the occurrence of ``superradiant" (i.e.~photon condensate) states in electron systems coupled to a spatially-varying electromagnetic field. 

In three spatial dimensions, the criterion, reported in Eq.~(\ref{eq:Stiffness7}), is {\it identical} to the Condon criterion for the occurrence of magnetic domains. 
The Zeeman coupling of the electronic spin degrees of freedom to the cavity field leads to the criterion in Eq.~(\ref{eq:EminFinal2}) and implies that in a real material one needs to know both orbital and spin non-local response functions to make quantitative predictions on the occurrence of a photon condensate phase.  

Finally, the condition for the occurrence of photon condensates in 2D systems embedded in quasi-2D cavities is reported in Eq.~(\ref{eq:2d_condition1}) and poses severe bounds on the observability of this phenomenon. We have indeed shown that in order to satisfy this criterion in the quasi-homogeneous limit, one needs to hunt for materials with a divergent orbital paramagnetic character. A few possibilities have been discussed in Sect.~\ref{sect:2D_discussion}.

While we have made no assumptions on the electromagnetic field, we have taken the electron system at hand to be homogeneous, i.e.~we have worked with the so-called ``jellium model"~\cite{Giuliani_and_Vignale}. Furthermore, relativistic Hamiltonian terms, such as spin-orbit coupling, have been neglected. In the future we plan to extend our investigations of photon condensate states to more general model Hamiltonians, especially ones that transcend the assumption of homogeneity.

The prediction of the possible coexistence in strongly correlated materials of exotic orders and photon condensate states requires accurate microscopic theories of the non-local orbital and spin response functions that take into account the role of electron-electron interactions.

\acknowledgements
We thank Pierre Nataf and Denis Basko for useful discussions.

This work was partially supported by the European Union's Horizon 2020 research and innovation programme under grant agreements No.~785219 - GrapheneCore2 and No.~881603 - GrapheneCore3.


 Work in Austin was supported by the Army Research Office (ARO) Grant \# W911NF-17-1-0312 (MURI).
Work in Catania was supported by the Universit\`a degli Studi di Catania,  Piano di Incentivi per la Ricerca di Ateneo 2020/2022, progetto Q-ICT.

As this manuscript was being finalized for publication, we learned about related work by Guerci et al.~\cite{guerci}, where a similar criterion for the occurrence of a superradiant phase transition in a cavity with a single mode was obtained. It is a great pleasure to thank Daniele Guerci, Pascal Simon, and Christophe Mora for sharing their results with us prior to publication.

This article is dedicated to the memory of Federico Tonielli.

\appendix

\onecolumngrid

\section{Disentangling light and matter}
\label{appendix:disentagling}
In this Appendix, we show that, in the thermodynamic $N\to \infty$, $V\to \infty$ limit (with $N/V = {\rm constant}$), it is permissible to assume a factorized ground state of the form
\begin{equation}
\label{eq:Factorization}
\ket{\Psi}=\ket{\psi}\ket{\Phi}~.
\end{equation}

We begin by defining the electron-photon Hamiltonian $\hat{\cal H}_{\rm el-ph}\equiv\hat{\cal H}^{(1)}_{\rm el-ph}+\hat{\cal H}^{(2)}_{\rm el-ph}$, where
\begin{equation}
\label{eq:Hamiltonians2}
\hat{\cal H}^{(1)}_{\rm el-ph} \equiv \sum_{i=1}^{N} \frac{e}{ m c}   \hat{{\bm A}}( \hat{\bm r}_i) \cdot \hat{ {\bm p}}_i
\end{equation}
and
\begin{equation}
\label{eq:Hamiltonians3}
\hat{\cal H}^{(2)}_{\rm el-ph}\equiv\sum_{i=1}^{N} \frac{e^2 }{2mc^2}\hat{\bm A}^2( \hat{\bm r}_i)~.
\end{equation}

The photon Hamiltonian $\hat{\cal H}_{\rm ph}$ has been defined in the main text. Let us split the matter Hamiltonian into the sum of kinetic and potential terms, i.e.~we write  $\hat{\cal H}\equiv\hat{\cal H}_{\rm K}+\hat{\cal H}_{\rm V}$, where:
\begin{equation}
\label{eq:Hamiltonians4}
\hat{\cal H}_{\rm K}\equiv\sum_{i=1}^N  \frac{\hat{\bm p}^2_i}{2m}
\end{equation}
and
\begin{equation}
\label{eq:Hamiltonians5}
\hat{\cal H}_{\rm V} \equiv  \frac{1}{2}\sum_{i\neq j} v(\hat{\bm r}_{i} - \hat{\bm r}_{j})~.
\end{equation}
In order to guarantee the correct thermodynamic limit, $\hat{\cal H}_{\rm el-ph}$, $\hat{\cal H}_{\rm ph}$, and $\hat{\cal H}$ must scale {\it extensively} with $N$. This implies that photonic and electronic operators must scale properly with $N$ in the $N \to \infty$ limit.  Let us discuss this fact explicitly. 

We begin by considering the photon Hamiltonian  $\hat{\cal H}_{\rm ph}$. We denote by the symbol $N_{\rm modes}$ the number of ``non-negligible" modes, i.e. modes that cannot be neglected in the thermodynamic limit. The photon Hamiltonian $\hat{\cal H}_{\rm ph}$ can have an extensive scaling with $N$ in two different cases: 
\begin{itemize}
\item $N_{\rm modes}$  is an {\it intensive} quantity (i.e. $N_{\rm modes}$ does not scale with $N$). In this case, the operator $\hat{a}_{\bm{q}_0,\sigma}$ characterized by a given $\bm{q}_0$ acquires a macroscopic occupation $\hat{a}_{\bm{q}_0,\sigma}\sim \sqrt{N}$;
\item  $N_{\rm modes}$  is an {\it extensive} quantity, while the occupation number $\hat{a}^\dagger_{\bm{q}_0,\sigma}\hat{a}_{\bm{q}_0,\sigma}$ of each mode is not macroscopic, i.e.~  
$\hat{a}_{\bm{q},\sigma}\sim \sqrt{N/N_{\rm modes}} \sim 1$. We now show that this case is not relevant for the occurrence of photon condensation. The paramagnetic electron-photon interaction $\hat{\cal H}^{(1)}_{\rm el-ph} $ scales like:
\begin{eqnarray}
\label{eq:HPara}
\hat{\cal H}^{(1)}_{\rm el-ph} &\sim & \sum_{\bm{q}} A_{\bm{q}} \hat{a}^\dagger_{\bm{q},\sigma} \hat{\bm{j}}_{\rm{p}}(\bm{q}) ~.
\end{eqnarray}
In the case of interest,  $A_{\bm{q}} \hat{a}^\dagger_{\bm{q},\sigma}\sim 1/\sqrt{N_{\rm modes}}\sim 1/\sqrt{N}$, while $\sum_{\bm{q}}\hat{\bm{j}}_{\rm{p}}(\bm{q})$  is {\it extensive} in $N$. We therefore get the following scaling with $N$ of the paramagnetic contribution: $\hat{\cal H}^{(1)}_{\rm el-ph}\sim N/\sqrt{N_{\rm modes}}\sim \sqrt{N}$. In summary, if $N_{\rm modes}$ is extensive, we have $\hat{\cal H}^{(1)}_{\rm el-ph}/N \sim 1/\sqrt{N} \to 0$ in the limit $N\to \infty$. Since $\hat{\cal H}^{(1)}_{\rm el-ph}$ is responsible for lowering the energy of the photon condensate phase, the fact that it scales to zero in the thermodynamic limit excludes the possibility of a phase transition.
\end{itemize}

Since we are interested in photon condensation, from now on we will consider only the case in which a finite number of modes acquires a macroscopic occupation number, i.e.~we assume that $N_{\rm modes}$  is an {\it intensive} quantity.  In this case, Hamiltonians (\ref{eq:Hamiltonians2}) and (\ref{eq:Hamiltonians3}) are {\it extensive}. 
Let us now focus on electronic operators. Being a sum of $N$ independent terms, $\hat{\cal H}_{\rm K}$ in Eq.~\eqref{eq:Hamiltonians4} is explicitly  {\it extensive}. Conversely, $\hat{\cal H}_{\rm V}$ in Eq.~\eqref{eq:Hamiltonians5} contains  a double sum, and is therefore expected to scale like $N^2$. Nevertheless, it is possible to show that, due to the ground-state equilibrium condition (i.e.~charge neutrality~\cite{Giuliani_and_Vignale}),  the expectation value of $\hat{\cal H}_{\rm V}$ over the equilibrium ground-state $|\psi\rangle$ scales with $N$. Below, we will therefore work with the rescaled operators $\hat{\cal H}/N$, $\hat{\cal H}_{\rm ph}/N$, and $\hat{\cal H}_{\rm el-ph}/N$, which are well defined in the thermodynamic $N\to \infty$ limit.

In order to prove Eq.~\eqref{eq:Factorization} we will show that in the limit $N\to \infty$
\begin{eqnarray}
\label{eq:HamiltoniansComm}
[\frac{\hat{\cal H}}{N},\frac{\hat{\cal H}_{\rm el-ph}}{N}]\to 0~,
\end{eqnarray}
and
\begin{eqnarray}
\label{eq:HamiltoniansComm1}
[\frac{\hat{\cal H}_{\rm ph}}{N},\frac{\hat{\cal H}_{\rm el-ph}}{N}]\to 0~.
\end{eqnarray}

The left-hand side of Eq.~\eqref{eq:HamiltoniansComm} contains three contributions, which we now carefully examine: 
\begin{itemize}
\item[a)] The first contribution is
\begin{eqnarray}
\label{eq:HamiltoniansComm2}
[\frac{\hat{\cal H}_{\rm K}}{N},\frac{\hat{\cal H}^{(1)}_{\rm el-ph}}{N}]&=&\sum_{i=1}^N  \frac{e\hbar }{2cm^2N^2} \Big[ \hat{ {\bm p}}_i\cdot{ {\bm q}} - \sum_{\bm{q},\sigma }  A_{\bm{q}}\big(\hat{a}_{\bm{q},\sigma} e^{i\bm{q}\cdot\hat{ \bm{r}}_i} - \hat{a}_{\bm{q},\sigma}^\dagger  e^{-i\bm{q}\cdot\hat{ \bm{r}}_i} \big)  \bm{u}_{\bm{q},\sigma}\cdot \hat{ {\bm p}}_i  +   \nonumber \\
&+&          \sum_{\bm{q},\sigma } A_{\bm{q}}\big(\hat{a}_{\bm{q},\sigma} e^{i\bm{q}\cdot\hat{ \bm{r}}_i} - \hat{a}_{\bm{q},\sigma}^\dagger  e^{-i\bm{q}\cdot\hat{ \bm{r}}_i} \big)      \hat{ {\bm p}}_i\cdot{ {\bm q}}~ \bm{u}_{\bm{q},\sigma}\cdot \hat{ {\bm p}}_i  \Big]~. 
\end{eqnarray}
This commutator vanishes like $1/N$, since $\sum_{i=1}^N$ scales like $N$, while terms like $\sum_{\bm{q}}  A_{\bm{q}} \hat{a}_{\bm{q},\sigma}\sim \sqrt{N/V}$ are of order unity in the limit $N, V \to \infty$ with $N/V={\rm constant}$.

\item[b)] The second contribution is
\begin{eqnarray}
\label{eq:HamiltoniansComm3}
[\frac{\hat{\cal H}_{\rm V}}{N},\frac{\hat{\cal H}^{(1)}_{\rm el-ph}}{N}]= \frac{1}{N^2} [ \frac{1}{2}\sum_{i\neq j} v(\hat{\bm r}_{i} - \hat{\bm r}_{j}),\sum_{j=1}^{N} \frac{e}{m c} \hat{{\bm A}}(\hat{\bm{r}}_j)  \cdot \hat{ {\bm p}}_j ]~.
\end{eqnarray}
Using that $[ f(\hat{\bm r}_i), \hat{ {\bm p}}_j ]=\delta_{i,j} i\hbar  \nabla_{\hat{\bm r}_i} f(\hat{\bm r}_i)$ and introducing the Coulomb force $\hat{\bm F}^{\rm C}_{i,j}=-\nabla_{\hat{\bm r}_i} v(\hat{\bm r}_{i} - \hat{\bm r}_{j})/2$ we get:
\begin{eqnarray}
\label{eq:HamiltoniansComm4}
[\frac{\hat{\cal H}_{\rm V}}{N},\frac{\hat{\cal H}^{(1)}_{\rm el-ph}}{N}]=- \sum_{i=1}^{N} \frac{i\hbar e   \hat{{\bm A}}(\hat{\bm{r}}_i) }{mcN^2}\cdot   \sum_{j\neq i} \hat{\bm F}^{\rm C}_{i,j} ~.
\end{eqnarray}
The quantity $\hat{\bm F}^{\rm T}_i\equiv \sum_{j\neq i} \hat{\bm F}^{\rm C}_{i,j} $ is the total force acting on the $i$-th particle. Even though the double sum in Eq.~(\ref{eq:HamiltoniansComm4}) brings in a factor scaling like $N^2$ in the large-$N$ limit, the expectation value of the commutator in Eq.~(\ref{eq:HamiltoniansComm4}) vanishes like $1/N$ in the $N\to \infty$ limit. This is due to the aforementioned charge-neutrality condition, which imposes that the expectation value of $\sum_{i=1}^N\hat{\bm F}^{\rm T}_i$ over the matter ground state $|\psi\rangle$ scales like $N$ in the $N\to \infty$ limit.

\item[c)] The third contribution is
\begin{eqnarray}
\label{eq:HamiltoniansComm3b}
[\frac{\hat{\cal H}}{N},\frac{\hat{\cal H}^{(2)}_{\rm el-ph}}{N}] = [\frac{\hat{\cal H}_{\rm K}}{N},\frac{\hat{\cal H}^{(2)}_{\rm el-ph}}{N}]&=&\sum_{i=1}^N  \frac{e^2\hbar }{2m^2c^2N^2} \sum_{\bm{q},\sigma }  \Big\{ \hat{ {\bm p}}_i\cdot{ {\bm q}}   \hat{{\bm A}}(\hat{\bm{r}}_i) \cdot\bm{u}_{\bm{q},\sigma}   A_{\bm{q}}\big(\hat{a}_{\bm{q},\sigma} e^{i\bm{q}\cdot\hat{ \bm{r}}_i} - \hat{a}_{\bm{q},\sigma}^\dagger  e^{-i\bm{q}\cdot\hat{ \bm{r}}_i} \big)   +   \nonumber \\
&+& \sum_{\bm{q},\sigma }\hat{{\bm A}}(\hat{\bm{r}}_i) \cdot\bm{u}_{\bm{q},\sigma}   A_{\bm{q}}\big(\hat{a}_{\bm{q},\sigma} e^{i\bm{q}\cdot\hat{ \bm{r}}_i} - \hat{a}_{\bm{q},\sigma}^\dagger  e^{-i\bm{q}\cdot\hat{ \bm{r}}_i} \big)          \hat{ {\bm p}}_i\cdot{ {\bm q}}                 \Big \} ~.
\end{eqnarray}
Again, this quantity scales to zero like $1/N$, since the sum $\sum_{i=1}^N$ brings in a factor $N$, while terms like $\sum_{\bm{q}, \sigma}  A_{\bm{q}} \hat{a}_{\bm{q},\sigma}$ and $\hat{{\bm A}}(\hat{\bm{r}}_i)$ are of order unity with respect to $N$.
\end{itemize}

In order to prove Eq.~\eqref{eq:HamiltoniansComm1}, it is convenient to rewrite the light-matter interaction Hamiltonian in terms of the real-space paramagnetic current $\hat{\bm{j}}_{\rm p}(\bm{r})$ and density $\hat{n}(\bm{r})$ operators:
\begin{eqnarray}
\label{eq:density_Appendix}
\hat{n}(\bm{r})&=&\sum_{i=1}^N \delta(\hat{\bm{r}}_i-{\bm{r}})~,\\
\hat{\bm{j}}_{\rm p}(\bm{r})&=&\frac{1}{2m}\sum_{i=1}^N \big[\hat{ \bm{p}}_i\delta(\hat{\bm{r}}_i-{\bm{r}})+\delta(\hat{\bm{r}}_i-{\bm{r}})\hat{\bm{p}}_i  \big]~.
\end{eqnarray}
Exploiting these definitions, we can write Eqs.~\eqref{eq:Hamiltonians2}-\eqref{eq:Hamiltonians3} as:
\begin{eqnarray}
\label{eq:Hamiltonians}
\hat{\cal H}^{(1)}_{\rm el-ph}&= & \frac{e}{ c} \int d^3{\bm r}~\hat{\bm{j}}_{\rm p}(\bm{r})\cdot  \hat{{\bm A}}( {\bm{r}})  ~,\\
\hat{\cal H}^{(2)}_{\rm el-ph}&= &\frac{e^2 }{2m c^2} \int d^3\bm{r}~\hat{n}(\bm{r}) \hat{{\bm A}}^2( {\bm{r}})~.
\end{eqnarray}
Using the commutator $[\hat{a}_{\bm{q},\sigma},\hat{a}_{\bm{q}^\prime,\sigma^\prime}^\dagger]=\delta_{\bm{q},\bm{q}^\prime}\delta_{\sigma,\sigma^\prime}$, we can rewrite the left-hand side of Eq.~\eqref{eq:HamiltoniansComm1} as the sum of the following two terms:
\begin{eqnarray}
\label{eq:HamiltoniansComm6}
[\frac{\hat{\cal H}_{\rm ph}}{N},\frac{\hat{\cal H}^{(1)}_{\rm el-ph}}{N}]= \sum_{\bm{q},\sigma }  \frac{\hbar \omega_{\bm{q}}  }{N^2}\Big\{\frac{e}{c}  \int d^3\bm{r} ~ \hat{\bm{j}}_{\rm p}(\bm{r})\cdot \bm{u}_{{\bm{q}},\sigma}  A_{\bm{q}}\big(\hat{a}_{{\bm{q}},\sigma} e^{i\bm{q}\cdot \bm{r}} - \hat{a}_{{\bm{q}},\sigma}^\dagger  e^{-i\bm{q}\cdot \bm{r}} \big)\Big\}
\end{eqnarray}
and
\begin{eqnarray}
\label{eq:HamiltoniansComm7}
[\frac{\hat{\cal H}_{\rm ph}}{N},\frac{\hat{\cal H}^{(2)}_{\rm el-ph}}{N}]&=& \sum_{\bm{q},\sigma }  \frac{\hbar \omega_{\bm{q}}  }{N^2}
\Big\{ \frac{e}{2m c}  \int d^3\bm{r}~\hat{n}(\bm{r})\hat{{\bm A}}( {\bm{r}})
\cdot   \bm{u}_{{\bm{q}},\sigma}   A_{{\bm{q}},\sigma} \big( \hat{a}_{{\bm{q}},\sigma} e^{i\bm{q}\cdot \bm{r}} - \hat{a}_{{\bm{q}},\sigma}^\dagger  e^{-i\bm{q}\cdot \bm{r}} \big)\\    
&+& 
\frac{e}{2m c}  \int d^3\bm{r}~\hat{n}(\bm{r})   A_{{\bm{q}},\sigma} \big( \hat{a}_{{\bm{q}},\sigma} e^{i\bm{q}\cdot \bm{r}} - \hat{a}_{{\bm{q}},\sigma}^\dagger  e^{-i\bm{q}\cdot \bm{r}} \big)
\hat{{\bm A}}( {\bm{r}})
\cdot   \bm{u}_{{\bm{q}},\sigma} \Big \} ~.\nonumber
\end{eqnarray}
Again, both quantities scale like $1/N$, since $ \int d^3\bm{r}~\hat{n}(\bm{r})\sim N$ and $\int d^3\bm{r} ~ \hat{\bm{j}}_{\rm p}(\bm{r})\sim N$, while $\hat{\bm A}({\bm r})$ and $ \sum_{\bm{q},{\sigma}}$ do not scale with $N$ (since, as stated at the beginning of this Appendix, we are considering the situation in which $N_{\rm modes}$ does not scale with $N$).
\section{Disentangling light and matter in the Zeeman coupling case}
\label{appendix:disentagling2} 
In this Appendix, we show that, in the thermodynamic $N\to \infty$, $V\to \infty$ limit (with $N/V = {\rm constant}$), it is allowed to assume a factorized ground state of the form
\begin{equation}
\label{eq:FactorizationZ}
\ket{\Psi}=\ket{\psi}\ket{\Phi}~,
\end{equation}
also when a Zeeman-type electron-photon interaction is taken into account. In this case, the electron-photon Hamiltonian is given by
\begin{eqnarray}
\label{eq:Hamiltonians2Z}
\hat{\cal H}_{\rm el-ph} &\equiv & \frac{g\mu_{\rm B}}{2}\sum_{i=1}^N \hat{\bm{\sigma}}_i\cdot \hat{\bm{B}}(\bm{r}_i)~.
\end{eqnarray}
The electron Hamiltonian $\hat{\cal H}$ and the photon Hamiltonian $\hat{\cal H}_{\rm ph}$ have been
defined in the main text. 
We here report again the explicit form of the cavity magnetic field: $\hat{\bm{B}}(\bm{r})=\sum_{\bm{q},\sigma }  A_{\bm{q}}i {q}\bm{u}_{\rm{T},\bm{q},\sigma} \big( \hat{a}_{\bm{q} }e^{i\bm{q}\cdot \bm{r}} -     \hat{a}^\dagger_{\bm{q} }e^{-i\bm{q}\cdot \bm{r}}        \big)$.
Again, in order to assure thermodynamic consistency, we assume that a finite number of relevant modes (i.e.~a number that does not scale with $N$), parametrized by ${\bm q}_0$, acquires macroscopic occupation, i.e.~$\hat{a}_{\bm{q}_0,\sigma}\sim \sqrt{N}$.
Since the electron Hamiltonian does not depend on the spin operators $\hat{\bm{\sigma}}_i$, we have $[{\hat{\cal H}}/{N},{\hat{\cal H}_{\rm el-ph}}/{N}]=0$.

In order to prove Eq.~\eqref{eq:FactorizationZ} we only need to show that
\begin{eqnarray}
\label{eq:HamiltoniansComm1Z}
[\frac{\hat{\cal H}_{\rm ph}}{N},\frac{\hat{\cal H}_{\rm el-ph}}{N}]\to 0
\end{eqnarray}
in the $N\to \infty$ limit.

To this end, it is convenient to rewrite the electron-photon Hamiltonian $\hat{\cal H}_{\rm el-ph}$ as a function of the real-space spin density $\hat{\bm{S}}(\bm{r})$, which is defined as following:
\begin{eqnarray}
\label{eq:spin_density_Appendix}
\hat{\bm{S}}(\bm{r})&=&\sum_{i=1}^N\hat{\bm{\sigma}}_i \delta(\hat{\bm{r}}_i-{\bm{r}})~.
\end{eqnarray}
Using this definition, we can rewrite Eq.~\eqref{eq:Hamiltonians2Z} as
\begin{eqnarray}
\label{eq:Hamiltonians2Zb}
\hat{\cal H}_{\rm el-ph} &= &   \frac{g\mu_{\rm B}}{2}\int d^3\bm{r}~\hat{\bm{S}}(\bm{r})\cdot \hat{\bm{B}}(\bm{r}) ~.
\end{eqnarray}
Exploiting the bosonic commutator $[\hat{a}_{{\bm{q}},\sigma} ,\hat{a}^\dagger_{{\bm{q}^\prime},\sigma^\prime} ] =\delta_{\bm{q},\bm{q}^\prime}\delta_{\sigma,\sigma^\prime}$, we can rewrite the left-hand side of Eq.~\eqref{eq:HamiltoniansComm1Z} as
\begin{eqnarray}
\label{eq:HamiltoniansComm6Z}
[\frac{\hat{\cal H}_{\rm ph}}{N},\frac{\hat{\cal H}_{\rm el-ph}}{N}]=- \sum_{\bm{q},\sigma }  \frac{ i\hbar q \omega_{\bm{q}} g\mu_{\rm B }}{2 N^2}\Big\{ \int d^3\bm{r} ~ \hat{\bm{S}}(\bm{r})\cdot \bm{u}_{\rm{T},\bm{q},\sigma} A_{\bm{q}}\big(\hat{a}_{{\bm{q}},\sigma} e^{i\bm{q}\cdot \bm{r}} + \hat{a}_{{\bm{q}},\sigma}^\dagger  e^{-i\bm{q}\cdot \bm{r}} \big)            \Big\}~.
\end{eqnarray}
This quantity scales like $1/N$, since $\int d^3\bm{r}~\hat{\bm S}(\bm{r})\sim N$, while $ \sum_{\bm{q},{\sigma} }$ and $A_{\bm{q}}\hat{a}_{{\bm{q}},\sigma}$ are of order unity with respect to $N$.

\section{Proof of Eq.~\eqref{eq:Ealpha2D}}

\label{app:details_bog}

The Hamiltonian in Eq.~(\ref{eqHph2d}) is a quadratic form of the photonic fields.
We now carry out a suitable Bogoliubov  transformation, switching from the bosonic operators $\hat{a}_{\bm{q}_\parallel,1 , n_z}$ and $\hat{a}^\dagger_{-\bm{q}_\parallel, 1, n_z}$ with odd $n_{z}$ to new bosonic operators $\hat{b}_{\bm{q}_\parallel, 1, j}$ and $\hat{b}^\dagger_{-\bm{q}_\parallel, 1, j}$ with integer $j$. Bosonic operators $\hat{a}_{\bm{q}_\parallel,1 , n_z}$ and $\hat{a}^\dagger_{-\bm{q}_\parallel, 1, n_z}$ with even mode index $n_{z}$ are decoupled from matter degrees of freedom. The Bogoliubov transformation reads as following:
\begin{equation}\label{eq:b-aadag}
\hat{b}_{\bm{q}_\parallel,1 , j} = \sum_{\ell} [X_{j,\ell}({\bm q}_\parallel) \hat{a}_{\bm{q}_\parallel, 1, 2 \ell+1} + 
Y_{j,\ell}({\bm q}_\parallel) \hat{a}^\dagger_{-\bm{q}_\parallel, 1, 2 \ell+1}]~,
\end{equation}
with $\ell,j$ integers.
Applying the Hermitian conjugation to the expression above and replacing $\bm{q}_\parallel \to -\bm{q}_\parallel$, one has
\begin{eqnarray}\label{eq:bdag-aadag}
\hat{b}^\dagger_{-\bm{q}_\parallel,1 , j} = \sum_{\ell}  
[Y^\ast_{j,\ell}(-{\bm q}_\parallel) \hat{a}_{\bm{q}_\parallel, 1, 2 \ell+1}
X^\ast_{j,\ell}(-{\bm q}_\parallel) \hat{a}^\dagger_{\bm{-q}_\parallel, 1, 2 \ell+1}]
~. 
\end{eqnarray}
For every ${\bm q}_\parallel$, we can therefore write the Bogoliubov transformation in the following compact form
\begin{equation}
\begin{bmatrix}
\{\hat{b}_{\bm{q}_\parallel,1 , j}\}\\
\{\hat{b}^\dagger_{-\bm{q}_\parallel, 1, j}\}
\end{bmatrix} 
=
\begin{bmatrix}
X({\bm q}_\parallel) && Y({\bm q}_\parallel)\\
Y^\ast({-\bm q}_\parallel) && X^\ast({-\bm q}_\parallel)
\end{bmatrix} 
\begin{bmatrix}
\{\hat{a}_{\bm{q}_\parallel, 1, 2 \ell+1}\}\\
\{\hat{a}^\dagger_{-\bm{q}_\parallel, 1,  2 \ell+1 }\}
\end{bmatrix}
~.
\end{equation}
It acts only on the photon modes with odd mode index and it is independent of the direction of the polarization vector ${\bm u}_{{\bm q}_\parallel, 1}$ . 
For this reason,  we have omitted the polarization label $\sigma=1$  from the Bogoliubov transformation matrices 
$X({\bm q}_\parallel)$ and $Y({\bm q}_\parallel)$.

We would like to find $X({\bm q}_\parallel)$ and $Y({\bm q}_\parallel)$ such that:
\begin{eqnarray}\label{eq:Bog_1}
\hat{\cal H}_{\rm ph} +  \hat{\cal H}_{\rm d}=
\sum_{{\bm q}_\parallel}\left[ \sum_{{\rm even}~ n_z} \hbar \omega_{{\bm q}_\parallel,n_z} \left( \hat{a}^\dagger_{\bm{q}_\parallel,1,n_z} \hat{a}_{\bm{q}_\parallel,1, n_z}+\frac{1}{2}\right)
+ \sum_j \hbar \Omega_{{\bm q}_\parallel,j} \left( \hat{b}^\dagger_{\bm{q}_\parallel,1,j} \hat{b}_{\bm{q}_\parallel,1, j}+\frac{1}{2}\right)\right]~,
\end{eqnarray}
with a suitable choice of $\Omega_{{\bm q}_\parallel,j}$. Notice that, differently from the main text, we have restored the vacuum contribution.
If (\ref{eq:Bog_1}) holds true, one has
\begin{equation}\label{eq:Bog_2}
[\hat{\cal H}_{\rm ph} +  \hat{\cal H}_{\rm d},\hat{b}_{\bm{q}_\parallel,1, j}]=- \hbar \Omega_{{\bm q}_\parallel,j}  \hat{b}_{\bm{q}_\parallel,1, j}~.
\end{equation}
Using Eq.~(\ref{eq:b-aadag}) we can write Eq.~(\ref{eq:Bog_2}) as
\begin{eqnarray}
&&\sum_{\ell} [ \hat{\cal H}_{\rm ph} +  \hat{\cal H}_{\rm d},  X_{j,\ell}({\bm q}_\parallel) \hat{a}_{\bm{q}_\parallel, 1, 2 \ell+1} + 
Y_{j,\ell}({\bm q}_\parallel) \hat{a}^\dagger_{-\bm{q}_\parallel, 1, 2 \ell+1} ] \nonumber\\
&=& - \hbar \Omega_{{\bm q}_\parallel,j}  \sum_{\ell} X_{j,\ell}({\bm q}_\parallel) \hat{a}_{\bm{q}_\parallel, 1, 2 \ell+1} + 
Y_{j,\ell}({\bm q}_\parallel) \hat{a}^\dagger_{-\bm{q}_\parallel, 1, 2 \ell+1}~,\nonumber\\ 
\end{eqnarray}
which is equivalent to
\begin{eqnarray}\label{eq:bog_2d}
&& \hbar \Omega_{{\bm q}_\parallel,j}  \sum_{\ell} X_{j,\ell}({\bm q}_\parallel) \hat{a}_{\bm{q}_\parallel, 1, 2 \ell+1} + 
Y_{j,\ell}({\bm q}_\parallel) \hat{a}^\dagger_{-\bm{q}_\parallel, 1, 2 \ell+1} \nonumber\\
&=&
\sum_{k} X_{j k}({\bm q}_\parallel) [\hbar \omega_{{\bm q}_\parallel,2k+1} 
\hat{a}_{{\bm q}_\parallel,1,2k+1}
+\frac{N}{m}\sum_\ell g_{k}(\bm{q}_\parallel) g_\ell(\bm{q}_\parallel) (\hat{a}_{{\bm q}_\parallel,1,2\ell+1}+\hat{a}^\dagger_{-{\bm q}_\parallel,1,2\ell+1})] \nonumber\\
&-& Y_{j k}({\bm q}_\parallel)[\hbar \omega_{{\bm q}_\parallel,2\ell+1} \hat{a}^\dagger_{{\bm q}_\parallel,1,2k+1}
+\frac{N}{m}\sum_\ell g_k(\bm{q}_\parallel) g_\ell(\bm{q}_\parallel) (\hat{a}_{{\bm q}_\parallel,1,2\ell+1}+\hat{a}^\dagger_{-{\bm q}_\parallel,1, 2\ell+1})]~,
\end{eqnarray}
where $g_j(\bm{q}_\parallel)=(-1)^{j}\sqrt{2D/\omega_{\bm{q}_\parallel,2j+1}}$.
The expression above can be written  compactly as
\begin{equation}\label{eq:eigenmodes}
\left({\cal K}_{{\bm q}_\parallel}-\hbar \Omega_{{\bm q}_\parallel,j} \openone_{2 N_{\rm max}}  \right) {\bm v}_j({\bm q}_\parallel)=0~,
\end{equation}
where we introduced a cutoff $N_{\rm max}$ on the number of modes in order to deal with finite-size matrices. The vector ${\bm v}_j({\bm q}_\parallel)$ reads as following:

\begin{eqnarray}
{\bm v}_j({\bm q}_\parallel)&=&[ X_{j, 0}({\bm q}_\parallel)  \hat{a}_{\bm{q}_\parallel, 1, 1} ,\cdots,X_{j, N_{\rm max}-1}({\bm q}_\parallel) \hat{a}_{\bm{q}_\parallel, 1, 2N_{\rm max}- 1},\nonumber \\
&& Y_{j, 0}({\bm q}_\parallel)  \hat{a}^\dagger_{-\bm{q}_\parallel, 1, 1} ,\cdots,Y_{j, N_{\rm max}-1}({\bm q}_\parallel) 
\hat{a}^\dagger_{-\bm{q}_\parallel, 1, 2N_{\rm max}- 1} ]^{\top}~.\nonumber \\
\end{eqnarray}

The solutions  of the linear-algebra problem posed by Eq.~(\ref{eq:eigenmodes})  can be found by setting to zero the determinant of the matrix ${\cal K}_{{\bm q}_\parallel}-\hbar \Omega_{{\bm q}_\parallel,j} \openone_{2 N_{\rm max}}$:
\begin{equation}\label{eq:determinantK}
{\rm Det}[{\cal K}_{{\bm q}_\parallel}-\hbar \Omega_{{\bm q}_\parallel,j} \openone_{2 N_{\rm max}}  ] =0~.
\end{equation}
The calculation of this determinant is a purely mathematical issue and is postponed to Appendix~\ref{app:calculation_determinant}. The final result is reported in Eq.~(\ref{eq:Detfinal}). Using this result and taking the $N_{\rm max}\to \infty$ limit, we find that the eigenvalues of the matrix ${\cal K}_{{\bm q}_\parallel}$ are the roots of the following transcendental equation:
\begin{equation}
1+\frac{n_{\rm 2D}}{m}\frac{2 \pi e^2 }{ c^2}
\frac{\tan\left(L_z\sqrt{\epsilon_{\rm r} { \Omega_{{\bm q}_\parallel,j}}^2/c^2-q_\parallel^2}/2\right)}{\sqrt{\epsilon_{\rm r} { \Omega_{{\bm q}_\parallel,j} }^2/c^2-q_\parallel^2}}=0~,
\end{equation}
where $n_{\rm 2D}=N/S$. 
Since ${\cal K}_{{\bm q}_\parallel}={\cal K}_{-{\bm q}_\parallel}$, one has $\Omega_{{\bm q}_\parallel,j}=\Omega_{-{\bm q}_\parallel,j}$
and ${\bm v}_j({\bm q}_\parallel)={\bm v}_j(-{\bm q}_\parallel)$, i.e.~$X({\bm q}_\parallel)=X(-{\bm q}_\parallel)$ and $ Y({\bm q}_\parallel)=Y(-{\bm q}_\parallel)$.

Similarly to what done above, we now calculate the following commutator:
\begin{equation}
[\hat{\cal H}_{\rm ph} +  \hat{\cal H}_{\rm d},\hat{b}^\dagger_{-\bm{q}_\parallel,1, j}]= \hbar \Omega_{-{\bm q}_\parallel,j}  \hat{b}^\dagger_{-\bm{q}_\parallel,1, j}~.
\end{equation}
Using Eq.~(\ref{eq:bdag-aadag}), we find
\begin{eqnarray}\label{eq:bog_2d_dag}
&&\hbar \Omega_{-{\bm q}_\parallel,j} \sum_{\ell}   
Y^\ast_{j,\ell}(-{\bm q}_\parallel) \hat{a}_{\bm{q}_\parallel, 1, 2 \ell+1}+
X^\ast_{j,\ell}(-{\bm q}_\parallel) \hat{a}^\dagger_{\bm{-q}_\parallel, 1, 2 \ell+1} \nonumber\\ 
&=&\sum_{k} X^\ast_{j k}(-{\bm q}_\parallel) [\hbar \omega_{{\bm q}_\parallel,2k+1} \hat{a}^\dagger_{-{\bm q}_\parallel,1,2k+1}
+\frac{N}{m}\sum_\ell g_k(\bm{q}_\parallel) g_\ell(\bm{q}_\parallel) (\hat{a}_{{\bm q}_\parallel,1,2\ell+1}
+\hat{a}^\dagger_{-{\bm q}_\parallel,1,2\ell+1})]  \nonumber\\
&-&Y_{j k}^\ast(-{\bm q}_\parallel)[\hbar \omega_{{\bm q}_\parallel,2\ell+1} \hat{a}_{{\bm q}_\parallel,1,2k+1} +\frac{N}{m}\sum_\ell g_k(\bm{q}_\parallel) g_\ell(\bm{q}_\parallel) (\hat{a}_{{\bm q}_\parallel,1,2\ell+1}+\hat{a}^\dagger_{-{\bm q}_\parallel,1, 2\ell+1})]~.
\end{eqnarray}
The expression above can be written as
\begin{equation}\label{eq:eigenmodess}
[ {\cal K}_{{\bm q}_\parallel}-\hbar \Omega_{-{\bm q}_\parallel,j} \openone_{2 N_{\rm max}}  ] {\bm v}^\ast_j(-{\bm q}_\parallel)=0~,
\end{equation}
where $\Omega_{-{\bm q}_\parallel,j}=\Omega_{{\bm q}_\parallel,j}$.
Since this eigenvalue problem is identical to Eq.~(\ref{eq:eigenmodes}), one has ${\bm v}^\ast_j(-{\bm q}_\parallel)={\bm v}_j({\bm q}_\parallel)$, i.e.~$X({\bm q}_\parallel)=X^\ast(-{\bm q}_\parallel)$ and $ Y({\bm q}_\parallel)=Y^\ast(-{\bm q}_\parallel)$.

Because of the properties of the matrices $X({\bm q}_\parallel)$ and $Y({\bm q}_\parallel)$, i.e.~$X({\bm q}_\parallel)=X^\ast(-{\bm q}_\parallel)=
X(-{\bm q}_\parallel)$ and $Y({\bm q}_\parallel)=Y^\ast(-{\bm q}_\parallel)=Y(-{\bm q}_\parallel)$, we can write
\begin{equation}
\begin{bmatrix}
\{\hat{b}_{\bm{q}_\parallel,1 , j}\}\\
\{\hat{b}^\dagger_{-\bm{q}_\parallel, 1, j}\}
\end{bmatrix} 
 =
 \begin{bmatrix}
 X({\bm q}_\parallel) && Y({\bm q}_\parallel)\\
 Y({\bm q}_\parallel) && X({\bm q}_\parallel)
 \end{bmatrix} 
\begin{bmatrix}
\{\hat{a}_{\bm{q}_\parallel, 1, 2 \ell+1}\}\\
\{\hat{a}^\dagger_{-\bm{q}_\parallel, 1,  2 \ell+1 }\}
\end{bmatrix}~.
 \end{equation}
Imposing the bosonic commutation rules, $[\hat{b}_{\bm{q}_\parallel,1 , j},\hat{b}^\dagger_{\bm{q}^\prime_\parallel,1 , j^\prime}]=\delta_{\bm{q}_\parallel,\bm{q}^\prime_\parallel}\delta_{j,j^\prime}$ and $[\hat{b}_{\bm{q}_\parallel,1 , j},\hat{b}_{\bm{q}^\prime_\parallel,1 , j^\prime}]=0$, we obtain the following properties
\begin{equation}
X({\bm q}_\parallel) X^{\top}({\bm q}_\parallel)- Y({\bm q}_\parallel) Y^{\top}({\bm q}_\parallel)= \openone~,
\end{equation} 
and 
\begin{equation}
X({\bm q}_\parallel) Y^\top({\bm q}_\parallel)- Y({\bm q}_\parallel) X^\top({\bm q}_\parallel)= 0~.
\end{equation}
By using the properties above, it is easy to obtain the inverse Bogoliubov transformation
 \begin{equation}
\begin{bmatrix}
\{\hat{a}_{\bm{q}_\parallel, 1, 2 \ell+1}\}\\
\{\hat{a}^\dagger_{-\bm{q}_\parallel, 1,  2 \ell+1 }\}
\end{bmatrix}
 =
 \begin{bmatrix}
 X^\top({\bm q}_\parallel) && -Y^\top({\bm q}_\parallel)\\
 -Y^\top({\bm q}_\parallel) && X^\top({\bm q}_\parallel)
 \end{bmatrix} 
\begin{bmatrix}
\{\hat{b}_{\bm{q}_\parallel,1 , j}\}\\
\{\hat{b}^\dagger_{-\bm{q}_\parallel, 1, j}\}
\end{bmatrix}  
 ~.
 \end{equation}
In terms of the new bosonic operators $\hat{b}_{ \bm{q}_\parallel,1,j} ^\dagger, \hat{b}_{ \bm{q}_\parallel,1,j}$, the effective Hamiltonian reads as following:
\begin{eqnarray}
&\hat{\cal{H}}^{\rm eff}_{\rm ph}&[{\psi}]=\braket{ \psi|\hat{\cal H}_{\rm 2D}|\psi} +
\sum_{{\bm q}_\parallel}\left\{ \sum_{{\rm even}~ n_z} \hbar \omega_{{\bm q}_\parallel,n_z}  \left(\hat{a}^\dagger_{\bm{q}_\parallel,1,n_z} \hat{a}_{\bm{q}_\parallel,1, n_z}+\frac{1}{2}\right)\right.
\nonumber\\
&+& \left.\sum_j \hbar \Omega_{{\bm q}_\parallel,j}\left( \hat{b}^\dagger_{\bm{q}_\parallel,1,j} \hat{b}_{\bm{q}_\parallel,1, j}+\frac{1}{2}\right)
+ {\cal J}_{{\bm q}_\parallel,1} \sum_{j,\ell} g_\ell(\bm{q}_\parallel) \left( \hat{b}^\dagger_{\bm{q}_\parallel,1, j} + \hat{b}_{-\bm{q}_\parallel,1, j} \right)   \left[X_{ j \ell }({\bm q}_\parallel)-Y_{j \ell}({\bm q}_\parallel)\right]
\right\}~.
\end{eqnarray}
The previous Hamiltonian can be written in a form that is manifestly Hermitian:
\begin{eqnarray}
&\hat{\cal{H}}^{\rm eff}_{\rm ph}&[{\psi}]=\braket{ \psi|\hat{\cal H}_{\rm 2D}|\psi} +
\sum_{{\bm q}_\parallel,1}\left\{ \sum_{{\rm even}~ n_z} \hbar \omega_{{\bm q}_\parallel,n_z}\left( \hat{a}^\dagger_{\bm{q}_\parallel,1,n_z} \hat{a}_{\bm{q}_\parallel,1, n_z}+\frac{1}{2}\right)\right.
\nonumber\\
&+& \left.\sum_j \hbar \Omega_{{\bm q}_\parallel,j}\left( \hat{b}^\dagger_{\bm{q}_\parallel,1,j} \hat{b}_{\bm{q}_\parallel,1, j}+\frac{1}{2}\right)
+ \left[{\cal J}_{{\bm q}_\parallel,1} \sum_{j,\ell} g_\ell(\bm{q}_\parallel)  \hat{b}^\dagger_{\bm{q}_\parallel,1, j}   \left[X_{ j \ell }({\bm q}_\parallel)-Y_{j \ell}({\bm q}_\parallel)\right]  +\rm{H.c.}\right]\right\}~.
\end{eqnarray}
In the effective Hamiltonian above, the even photon modes are independent of the light-matter interaction, while the odd photon modes are renormalized by the diamagnetic term and expressed as
a sum of displaced harmonic oscillators. 
For every matter state $|\psi\rangle$, the ground state $\ket{\Phi}$ of $\hat{\cal{H}}^{\rm eff}_{\rm ph}[\psi]$ is therefore a tensor product $\ket{\mathscr{B}} \equiv \otimes_{\bm{q}_\parallel,j} \ket{\beta_{\bm{q}_\parallel,1,j}}$ 
of coherent states of the $\hat{b}_{\bm{q}_\parallel,1,j}$ operators, i.e.~$\hat{b}_{\bm{q}_\parallel^\prime,1,\ell} \ket{\mathscr{B}}=\beta_{\bm{q}^\prime_\parallel,1,\ell}   \ket{\mathscr{B}}$. 

We now introduce the following energy functional, 
obtained by taking the expectation value of $\hat{\cal{H}}^{\rm eff}_{\rm ph}[\psi]$ over $ \ket{\mathscr{B}}$: 
$E[\{\beta_{\bm{q}_\parallel, 1,j}\}, \psi]\equiv \braket{\Psi|\hat{\cal H}_{\bm A}|\Psi}=\braket{\mathscr{B}| \hat{\cal{H}}^{\rm eff}_{\rm ph}[\psi]|\mathscr{B}}$:
\begin{eqnarray}
E[\{\beta_{\bm{q}_\parallel, 1,j}\}, \psi]&=&
\braket{\psi|{\cal H}_{\rm 2D} |\psi}+
 \sum_{{\bm q}_\parallel,j} 
\Big[ 
 \hbar \Omega_{{\bm q}_\parallel,j}\big( \beta^\ast_{\bm{q}_\parallel,1, j} \beta_{\bm{q}_\parallel,1, j}+\frac{1}{2}\big)
\nonumber \\
&+& {\cal J}({\bm q}_\parallel,1) 
(   \beta^\ast_{\bm{q}_\parallel,1, j} +\beta_{-\bm{q}_\parallel,1, j}  ) \sum_{\ell} g_\ell(\bm{q}_\parallel) 
 (X_{j \ell }({\bm q}_\parallel)-Y_{j \ell}({\bm q}_\parallel)) 
\Big]
~.
\end{eqnarray}
We now observe that the order parameter ${\alpha}_{{\bm q}_\parallel,1,2 \ell+1}$ introduced in the main text is linearly-dependent on ${\beta}_{{\bm q}_\parallel,1,j}$, i.e.
\begin{equation}\label{eq:alphabeta}
\begin{bmatrix}
\{\alpha_{\bm{q}_\parallel, 1,2\ell+1}\}\\
\{\alpha^\ast_{-\bm{q}_\parallel, 1, 2\ell+1}\}
\end{bmatrix} 
=
\begin{bmatrix}
X^\top({\bm q}_\parallel) && -Y^\top({\bm q}_\parallel)\\
-Y^\top({\bm q}_\parallel) && X^\top({\bm q}_\parallel)
\end{bmatrix} 
\begin{bmatrix}
\{\beta_{\bm{q}_\parallel,1, j}\}\\
\{\beta^\ast_{-\bm{q}_\parallel,1, j}\}
\end{bmatrix}~.
\end{equation}
By using the linear relation above, we can express the energy functional $E[\{\beta_{\bm{q}_\parallel, 1,j}\}, \psi]$ in terms of $\{\alpha_{\bm{q}_\parallel, 1,j}\}$. Carrying out such procedure and neglecting the vacuum energy, we finally obtain Eq. \eqref{eq:Ealpha2D} of the main text.

\section{Calculation of the determinant in Eq.~\eqref{eq:determinantK}}
\label{app:calculation_determinant}

In this Appendix we calculate the determinant in the left-hand side of Eq.~\eqref{eq:determinantK}. To this end, it is useful to write the matrix ${\cal K}_{{\bm q}_\parallel}-\hbar  \Omega_{{\bm q}_\parallel,j} \openone_{2 N_{\rm max}} $ defined in  Eq. \eqref{eq:eigenmodes} in the following block form:
\begin{equation}
{\cal K}_{{\bm q}_\parallel}-\hbar  \Omega_{{\bm q}_\parallel,j} \openone_{2 N_{\rm max}} =
\begin{bmatrix}
Q({\bm q}_\parallel) +V({\bm q}_\parallel) -\hbar  \Omega_{{\bm q}_\parallel,j} \openone_{N_{\rm max}} & -V({\bm q}_\parallel) \\
V({\bm q}_\parallel) & -Q({\bm q}_\parallel) -V({\bm q}_\parallel) - \hbar  \Omega_{{\bm q}_\parallel,j} \openone_{N_{\rm max}}  
\end{bmatrix}~,
\end{equation}
where
\begin{equation}\label{eq:Qmat}
Q_{k,\ell}({\bm q}_\parallel) =\hbar \omega_{\bm{q}_\parallel,2 \ell+1} \delta_{k,\ell} 
\end{equation}
and
\begin{equation}
V_{k,\ell}({\bm q}_\parallel)  = \frac{N}{m} g_k(\bm{q}_\parallel) g_\ell(\bm{q}_\parallel)~.
\end{equation}
Carrying out simple algebraic manipulations, we find
\begin{equation}
{\cal K}_{{\bm q}_\parallel}-\hbar \Omega_{{\bm q}_\parallel,j}  \openone_{2 N_{\rm max}} =
[\openone_{2 N_{\rm max}}+{\cal W}({\bm q}_\parallel)]
\begin{bmatrix}
Q({\bm q}_\parallel) -\hbar  \Omega_{{\bm q}_\parallel,j} \openone_{N_{\rm max}} & 0 \\
0 & -Q({\bm q}_\parallel)  -\hbar \Omega_{{\bm q}_\parallel,j} \openone_{N_{\rm max}}  
\end{bmatrix}~,
\end{equation}
where
\begin{equation}
{\cal W}({\bm q}_\parallel)=
\begin{bmatrix}
{\cal W}_-({\bm q}_\parallel) & {\cal W}_+({\bm q}_\parallel) \\
{\cal W}_-({\bm q}_\parallel) & {\cal W}_+({\bm q}_\parallel)  
\end{bmatrix}
\end{equation}
and
\begin{equation}
{\cal W}_\pm({\bm q}_\parallel)  =    V({\bm q}_\parallel)  \Big(\pm\hbar \Omega_{{\bm q}_\parallel,j} \openone_{N_{\rm max}} + Q({\bm q}_\parallel)  \Big)^{-1}~.
\end{equation}
Using the expressions above, we can write the determinant at hand as
\begin{equation}
{\rm Det}[{\cal K}_{{\bm q}_\parallel}-\hbar \Omega_{{\bm q}_\parallel,j}  \openone_{2 N_{\rm max}}]=
 \prod_{\ell}[(\hbar  \Omega_{{\bm q}_\parallel,j} )^2-(\hbar \omega_{\bm{q}_\parallel,2\ell+1})^2] {\rm Det}[\openone_{2 N_{\rm max}}+{\cal W}({\bm q}_\parallel)]~.
\end{equation}
We now focus on ${\rm Det}[\openone_{2 N_{\rm max}}+{\cal W}({\bm q}_\parallel)]$ and use the following well-known algebraic property,
\begin{equation}\label{eq:DetTr}
{\rm Det}[\openone_{2 N_{\rm max}}+{\cal W}({\bm q}_\parallel)]=\exp\{{\rm Tr}[\ln(\openone_{2 N_{\rm max}}+{\cal W}({\bm q}_\parallel) ] \}~.
\end{equation}
The trace in the right-hand side of the previous equation can be written as
\begin{equation}\label{eq:LogTr}
 {\rm Tr}[\ln(\openone_{2 N_{\rm max}}+{\cal W}({\bm q}_\parallel)]=\sum^\infty_{j=1} \frac{(-1)^{j-1}}{j} {\rm Tr}[{\cal W}^j({\bm q}_\parallel)]~.
\end{equation}
For block matrices, the following property holds true:
\begin{equation}
{\rm Tr}
\left\{
\begin{bmatrix}
A & B \\
A& B 
\end{bmatrix}
\begin{bmatrix}
C & D \\
C& D 
\end{bmatrix}
\right\}={\rm Tr}\{(A+B)(C+D)\}~.
\end{equation}
We therefore have
\begin{equation}
{\rm Tr}[{\cal W}^j({\bm q}_\parallel)]={\rm Tr}\{[{\cal W}_+({\bm q}_\parallel) +{\cal W}_-({\bm q}_\parallel)]^j\}~.
\end{equation}
Furthermore, it is possible to show that
\begin{equation}
{\rm rank}[{\cal W}_+({\bm q}_\parallel) +{\cal W}_-({\bm q}_\parallel)]=1~.
\end{equation}
The previous property of the matrix ${\cal W}_+({\bm q}_\parallel) +{\cal W}_-({\bm q}_\parallel)$ can be proved by direct inspection, showing that all the columns of ${\cal W}_+({\bm q}_\parallel) +{\cal W}_-({\bm q}_\parallel)$ can be obtained, for example, by multiplying the first column for a suitable constant.

We therefore conclude that ${\cal W}_+({\bm q}_\parallel) +{\cal W}_-({\bm q}_\parallel) $ has only one non-zero eigenvalue. As a consequence, we find that
\begin{equation}
{\rm Tr}[{\cal W}^j({\bm q}_\parallel)]=
{\rm Tr}\{{\cal W}_+({\bm q}_\parallel) +{\cal W}_-({\bm q}_\parallel)]^j\}=
{\rm Tr}[({\cal W}_+({\bm q}_\parallel) +{\cal W}_-({\bm q}_\parallel)]^j=
{\rm Tr}[{\cal W}({\bm q}_\parallel)]^j~.
\end{equation}
Replacing this result in Eq.~(\ref{eq:LogTr}), we therefore find that
\begin{equation}
 {\rm Tr}\{\ln[\openone_{2 N_{\rm max}}+{\cal W}({\bm q}_\parallel)]\}
 =\sum^\infty_{j=1} \frac{(-1)^{j-1}}{j} {\rm Tr}[{\cal W}({\bm q}_\parallel)]^j=\ln(1+ {\rm Tr}[{\cal W}({\bm q}_\parallel)])~,
\end{equation}
where
\begin{equation}\label{eq:Trfinale}
{\rm Tr}[{\cal W}({\bm q}_\parallel)]= {\rm Tr}[{\cal W}_+({\bm q}_\parallel)+{\cal W}_+({\bm q}_\parallel)]=\sum^{N_{\rm max}-1}_{\ell=0}
\frac{2 \omega_{{\bm q}_\parallel,2\ell+1} N g_\ell^2({\bm q}_\parallel)}{m \hbar(\omega_{{\bm q}_\parallel,2 \ell+1}^2-\Omega^2)}
=
\sum^{N_{\rm max}-1}_{\ell=0}
\frac{4 D N}{m \hbar(\omega_{{\bm q}_\parallel,2 \ell+1}^2- \Omega_{{\bm q}_\parallel,j}^2)}~.
\end{equation}
Replacing Eq.~(\ref{eq:Trfinale}) in Eq.~(\ref{eq:DetTr}), we find
\begin{equation}
{\rm Det}[\openone_{2 N_{\rm max}}+{\cal W}_{{\bm q}_\parallel}] =1+\sum^{N_{\rm max}-1}_{\ell=0}
\frac{4 D N}{m \hbar(\omega_{{\bm q}_\parallel,2 \ell+1}^2- \Omega_{{\bm q}_\parallel,j}^2)}=
1+\frac{n_{\rm 2D}}{m}\frac{2 \pi e^2 }{ c^2}
\frac{\tan\left(L_z\sqrt{\epsilon_{\rm r} {\Omega_{{\bm q}_\parallel,j} }^2/c^2-q_\parallel^2}/2\right)}{\sqrt{\epsilon_{\rm r} { \Omega_{{\bm q}_\parallel,j}}^2/c^2-q_\parallel^2}}
~.
\end{equation}
In the last equality we have used that $\sum_{\ell}[(2\ell+1)^2-x^2]^{-1}=\pi \tan(\pi x/2)/(4x)$ and the limit $N_{\rm max}\to \infty$ has been taken.
In summary, the final desired result is:
\begin{equation}\label{eq:Detfinal}
{\rm Det}[{\cal K}_{{\bm q}_\parallel}-\hbar  \Omega_{{\bm q}_\parallel,j}  \openone_{2 N_{\rm max}}]=
 \prod_{\ell}[(\hbar  \Omega_{{\bm q}_\parallel,j} )^2-(\hbar \omega_{\bm{q}_\parallel,2\ell+1})^2] 
 \left[1+\frac{n_{\rm 2D}}{m}\frac{2 \pi e^2 }{ c^2}
\frac{\tan\left(L_z\sqrt{\epsilon_{\rm r} { \Omega_{{\bm q}_\parallel,j} }^2/c^2-q_\parallel^2}/2\right)}{\sqrt{\epsilon_{\rm r} { \Omega_{{\bm q}_\parallel,j} }^2/c^2-q_\parallel^2}}\right]~.
\end{equation}
\section{Calculation of the determinant in Eq.~({\ref{eq:determinant_2D_stiffness}})}
\label{app:determinant_2d}

In this Appendix we calculate the determinant of the matrix ${\cal M}_{{\bm q}_\parallel}$  defined in Eq.~\eqref{eq:calM}. 
To this end, it is useful to write ${\cal M}_{{\bm q}_\parallel}$ in the following block form:
 \begin{eqnarray}
&&{\cal M}_{{\bm q}_\parallel} =
\begin{bmatrix}
{\cal M}^{(x)}_{{\bm q}_\parallel}  &0 \\
0 & {\cal M}^{(y)}_{{\bm q}_\parallel}  
\end{bmatrix}~,
\end{eqnarray}
where 
\begin{equation}
{\cal M}^{(x)}_{{\bm q}_\parallel}  = Q ({{\bm q}_\parallel}) +U ({{\bm q}_\parallel})~,
\end{equation}
\begin{equation}
{\cal M}^{(y)}_{{\bm q}_\parallel}  = Q ({{\bm q}_\parallel})~,
\end{equation}
$Q ({{\bm q}_\parallel})$ has been defined in Eq.~(\ref{eq:Qmat}), and
\begin{equation}
U_{k,\ell}({\bm q}_\parallel)  = \frac{2S\chi^{\rm J}_{\rm T}(q_\parallel,0)}{m} g_k(\bm{q}_\parallel) g_\ell(\bm{q}_\parallel)~.
\end{equation}
By exploiting the block decomposition, the determinant of ${\cal M}_{{\bm q}_\parallel}$ can be expressed as
\begin{equation}
\Delta_{{\bm q}_\parallel}= {\rm Det}({\cal M}_{{\bm q}_\parallel})={\rm Det}({\cal M}^{(x)}_{{\bm q}_\parallel}){\rm Det}({\cal M}^{(y)}_{{\bm q}_\parallel})~,
\end{equation}
where
\begin{equation}
 {\rm Det}({\cal M}^{(y)}_{{\bm q}_\parallel})=\prod_{\ell=0}^{N_{\rm max}-1} \hbar \omega_{\bm{q}_\parallel,2 \ell+1}~.
\end{equation}
Carrying out simple algebraic manipulations, we find
\begin{equation}
{\cal M}^{(x)}_{{\bm q}_\parallel} =[\openone_{N_{\rm max}}+U ({{\bm q}_\parallel})Q^{-1} ({{\bm q}_\parallel})]Q ({{\bm q}_\parallel})~.
\end{equation}
The expression of ${\rm Det}[{\cal M}_{{\bm q}_\parallel}]$ can be further simplified as
\begin{equation}
 \Delta_{{\bm q}_\parallel}= {\rm Det}[\openone_{ N_{\rm max}}+  U ({{\bm q}_\parallel})Q^{-1} ({{\bm q}_\parallel}) ]\prod_{\ell=0}^{N_{\rm max}-1} [\hbar\omega_{\bm{q}_\parallel,2 \ell+1}]^2~. 
\end{equation}
It is easy to verify that
each column of matrix $U ({{\bm q}_\parallel})Q^{-1} ({{\bm q}_\parallel})$ can be obtained by multiplying the first column for a suitable constant: this implies that
${\rm rank}[U ({{\bm q}_\parallel})Q^{-1} ({{\bm q}_\parallel})]=1$. 
Using this property, and following the same procedure discussed in the Appendix~\ref{app:calculation_determinant}, we have
\begin{eqnarray}
{\rm Det}[\openone_{ N_{\rm max}}+  W_{{\bm q}_\parallel}] &=& 1+{\rm Tr}[ U({\bm q}_\parallel)Q^{-1}({\bm q}_\parallel) ]
 =1+ \frac{2S\chi^{\rm J}_{\rm T}(q_\parallel,0)}{m} \sum^{N_{\rm max}-1}_{\ell=0} \frac{g_\ell^2(\bm{q}_\parallel)}{\hbar \omega_{\bm{q}_\parallel,2 \ell+1}}\nonumber\\
 &=& 1+ \frac{4S\chi^{\rm J}_{\rm T}(q_\parallel,0) D}{m \hbar} \sum^{N_{\rm max}-1}_{\ell=0} \frac{1}{\omega_{\bm{q}_\parallel,2 \ell+1}^2}~.
\end{eqnarray}
Taking the limit $N_{\rm max}\to \infty$, we find
\begin{equation}
 {\rm Det}[\openone_{ N_{\rm max}}+  W_{{\bm q}_\parallel}]= 1+ \chi^{\rm J}_{\rm T}(q_\parallel,0) \frac{2 \pi e^2}{c^2 q_\parallel} 
\tanh\Big(\frac{q_\parallel L_z}{2}\Big)~.
\end{equation}
In summary, the determinant of the matrix ${\cal M}_{{\bm q}_\parallel}$ is given by
\begin{equation}
 \Delta_{{\bm q}_\parallel}=\left[  1+ \chi^{\rm J}_{\rm T}(q_\parallel,0) \frac{2 \pi e^2}{c^2 q_\parallel} 
\tanh\Big(\frac{q_\parallel L_z}{2}\Big)  \right] \prod_{\ell} [\hbar\omega_{\bm{q}_\parallel,2 \ell+1}]^2~,
\end{equation}
as in Eq.~({\ref{eq:determinant_2D_stiffness}}) of the main text.


\begin{thebibliography}{77}
\bibitem{jellium}
Strictly speaking, in the (unscreened) jellium model~\cite{Giuliani_and_Vignale} $v(r)=e^2/(\epsilon_{\rm r} r)$, where $\epsilon_{\rm r}$ is a suitable background dielectric constant which will be introduced below. Nowhere in our proof we will assume this specific form of $v(r)$.
\bibitem{footnote_l_matter}
For example, in zero magnetic field, $\ell_{\rm matter}$ is expected to be on the order of $1/k_{\rm F}$, where $k_{\rm F}$ is the Fermi wave number of the 2DES. Similarly, in a perpendicular magnetic field ${\bm B}= B\hat{\bm z}$, $\ell_{\rm matter} \sim \ell_{\rm B}$, where $\ell_{\rm B} = \sqrt{\hbar c/(eB)}$ is the magnetic length.
\end{thebibliography}
\end{document}